\documentclass[preprint2]{aastex}
\usepackage{natbib}
\usepackage{times}
\usepackage{float}
\usepackage{wrapfig}
\usepackage{float}
\usepackage{longtable}
\usepackage{graphicx}
\usepackage{verbatim}
\usepackage{amssymb}
\usepackage{amsbsy}

\def\Mode{2} 
\def\InsertAllPartOne{1} 

\usepackage{fancyhdr} 
\fancyfootoffset[R]{1cm}

\usepackage{seqsplit}

\newcommand{\nii}{[N\,{\sc ii}] }

\newcommand{\oiii}{[O\,{\sc iii}] }

\newcommand{\oi}{[O\,{\sc i}] }

\newcommand{\cii}{C\,{\sc ii}\rm}

\newcommand{\ie}{i.\,e.\,}
\newcommand{\etal}{et al$.$ }

\newcommand{\arcm}{\mbox{\ensuremath{.\mkern-4mu^\prime}}}  
\newcommand{\arcs}{\mbox{\ensuremath{.\!\!^{\prime\prime}}}}
\newcommand{\degs}{\ensuremath{^{\circ}}}
\newcommand{\mum}{$\mu$m}
\newcommand{\ts}{\thinspace}


\shorttitle{GOALS {\em Herschel} Image Atlas and Aperture Photometry} 
\shortauthors{Chu \etal 2017}
                                                                                
\begin{document}
\title{
	\ifnum\Mode=2 
	\vskip -0.5truein
	{
	}
	\fi
The Great Observatories All-Sky LIRG Survey: {\em Herschel} Image Atlas and Aperture Photometry\footnote{Based on {\em Herschel Space Observatory} observations. {\em Herschel} is an ESA space observatory with science instruments provided by European-led Principal
Investigator consortia and with important participation from NASA.}}
\author{
Jason K. Chu\altaffilmark{1},
D. B. Sanders\altaffilmark{1},
K. L. Larson\altaffilmark{1,2},
J. M. Mazzarella\altaffilmark{2},
J. H. Howell\altaffilmark{2},
T. D\'iaz-Santos\altaffilmark{2,3},
K. C. Xu\altaffilmark{4},
R. Paladini\altaffilmark{4},
B. Schulz\altaffilmark{4},
D. Shupe\altaffilmark{4},
P. Appleton\altaffilmark{4},
L. Armus\altaffilmark{2},
N. Billot\altaffilmark{5},
B. H. P. Chan\altaffilmark{2},
A. S. Evans\altaffilmark{6,7},
D. Fadda\altaffilmark{2},
D. T. Frayer\altaffilmark{7},
S. Haan\altaffilmark{7},
C. M. Ishida\altaffilmark{8},
K. Iwasawa\altaffilmark{9},
D.-C. Kim\altaffilmark{7},
S. Lord\altaffilmark{10},
E. Murphy\altaffilmark{7},
A. Petric\altaffilmark{11},
G. C. Privon\altaffilmark{12},
J. A. Surace\altaffilmark{13},
E. Treister\altaffilmark{12}
}
\altaffiltext{1}{Institute for Astronomy, University of Hawaii, 2680
Woodlawn Drive, Honolulu, HI 96822; jasonchu@ifa.hawaii.edu, sanders@ifa.hawaii.edu}
\altaffiltext{2}{Infrared Processing \& Analysis Center, MS 100-22, California Institute of Technology, Pasadena, CA 91125; bchan, jhhowell, klarson, lee, mazz@ipac.caltech.edu, fadda@discovery.saclay.cea.fr}
\altaffiltext{3}{Nucleo de Astronom\'ia de la Facultad de Ingenier\'ia, Universidad Diego Portales, Av. Ejercito Libertador 441, Santiago, Chile; tanio.diaz@mail.udp.cl}
\altaffiltext{4}{NASA Herschel Science Center, MS 100-22, California Institute of Technology,
Pasadena, CA 91125; apple, bschulz, cxu, paladini, shupe@ipac.caltech.edu}
\altaffiltext{5}{Observatoire de l'Universit\'e de Gen\'eve, 51 chemin des Maillettes, 1290 Versoix, Switzerland; billot@iram.es}
\altaffiltext{6}{Department of Astronomy, University of Virginia, Charlottesville, VA 22904-4325; aevans@virginia.edu}
\altaffiltext{7}{National Radio Astronomy Observatory, 
520 Edgemont Road, Charlottesville, VA 22903-2475; dfrayer, dkim, emurphy@nrao.edu, sebhaan@gmail.com}
\altaffiltext{8}{Department of Physics and Astronomy, University of Hawaii at Hilo, Hilo, HI, 96720; cishida@hawaii.edu}
\altaffiltext{9}{ICREA and Institut del Ci\`encies del Cosmos (ICC), Universitat de Barcelona (IEEC-UB), Mart\'i i Franqu\`es 1, 08028 Barcelona, Spain; kazushi.iwasawa@icc.ub.edu}
\altaffiltext{10}{SETI Institute; slord@seti.org}
\altaffiltext{11}{Canada France Hawaii Telescope Corp., Conc\'epcion, Chile; petric@cfht.hawaii.edu}
\altaffiltext{12}{Instituto de Astrof\'isica, Facultad de F\'isica, Pontificia Universidad Cat\'olica de Chile, Casilla 306, Santiago 22, Chile; gprivon, etreiste@astro.puc.cl}
\altaffiltext{13}{Spitzer Science Center, MS 314-6, California Institute of Technology, Pasadena, CA 91125; jason@ipac.caltech.edu}

\begin{abstract}

Far-infrared (FIR) images and photometry are presented for 201 Luminous and Ultraluminous Infrared Galaxies [LIRGs: log${\ts(L_{\rm IR}/L_\odot) = 11.00 - 11.99}$, ULIRGs: log${\ts(L_{\rm IR}/L_\odot) = 12.00 - 12.99}$], in the Great Observatories All-Sky LIRG Survey (GOALS) based on observations with the {\it Herschel Space Observatory} Photodetector Array Camera and Spectrometer (PACS) and the Spectral and Photometric Imaging Receiver (SPIRE) instruments.  The image atlas displays each GOALS target in the three PACS bands (70, 100, and 160~\micron) and the three SPIRE bands (250, 350, and 500~\micron), optimized to reveal structures at both high and low surface brightness levels, with images scaled to simplify comparison of structures in the same physical areas of {$\sim$$100\times100$~kpc$^2$}.  Flux densities of companion galaxies in merging systems are provided where possible, depending on their angular separation and the spatial resolution in each passband, along with integrated system fluxes (sum of components).  This dataset constitutes the imaging and photometric component of the GOALS {\it Herschel} OT1 observing program, and is complementary to atlases presented for the {\em Hubble Space Telescope} (Evans \etal 2017, in prep.), {\em Spitzer Space Telescope} (Mazzarella \etal 2017, in prep.), and {\em Chandra X-ray Observatory} (Iwasawa \etal 2011, 2017, in prep.).  Collectively these data will enable a wide range of detailed studies of AGN and starburst activity within the most luminous infrared galaxies in the local Universe.

\end{abstract}


\section{Introduction}

The Great Observatories All-Sky LIRG Survey \cite[GOALS,][]{2009PASP..121..559A}, combines both imaging and spectroscopic data for the complete sample of 201 Luminous Infrared Galaxies (LIRGs:  log${\ts(L_{\rm IR}/L_\odot)}> 11.0$) selected from the IRAS Revised Bright Galaxy Sample \citep[RBGS,][]{2003AJ....126.1607S}.  The full RBGS contains 629 objects, representing a complete sample of extragalactic sources with IRAS 60 \mum~flux density, $S_{\rm 60} > 5.24\ts$Jy, covering the entire sky above a Galactic latitude of $\left|b\right| > 5\degs$.   The median redshift of objects in the GOALS sample is $\langle z\rangle = 0.021$, with a maximum redshift of $z_{\rm max} =  0.0876$.  As the nearest and brightest 60 \mum~extragalactic objects, they represent a sample that is the most amenable for study at all wavelengths.  

The primary objective of the GOALS multi-wavelength survey is to fully characterize the diversity of properties observed in a large, statistically significant sample of the nearest LIRGs.  This allows us to probe the full range of phenomena such as normal star formation, starbursts, and active galactic nuclei (AGN) that power the observed far-infrared emission, as well as to better characterize the range of galaxy types (\ie normal disks, major and minor interactions/mergers, etc.) that are associated with the LIRG phase.   A secondary objective is to provide a data set that is ideally suited for comparison with LIRGs observed at high redshifts.

GOALS currently includes imaging and spectroscopy from the {\it Spitzer}, {\it Hubble}, {\it GALEX}, {\it Chandra}, {\it XMM-Newton}, and now {\it Herschel} space-borne observatories, along with complementary ground-based observations from ALMA, Keck, and other telescopes.  The GOALS project is described in more detail at \url{http://goals.ipac.caltech.edu/}.

Due to limitations in angular resolution, wavelength coverage, and sensitivity of pre-{\em Herschel} ({\em IRAS}, {\em ISO}, {\em Spitzer}, {\em AKARI}) far-infrared (FIR) data, the spatial distribution of FIR emission within the GOALS sources, and the total amount of gas and dust in these systems, are poorly determined.  The {\em Herschel} data will allow us for the first time to directly probe the critical FIR and submillimeter wavelength regime of these infrared luminous systems, enabling us to accurately determine the bolometric luminosities, infrared surface brightnesses, star formation rates, and dust masses and temperatures on spatial scales of 2 -- 5 kpc within the GOALS sample.  

This paper presents imaging and photometry for all 201 LIRGs and LIRG systems in the IRAS RBGS that were observed during our GOALS {\it Herschel} OT1 program.  A more complete description of the GOALS sample is given in \S 2.  The data acquisition is described in \S 3 and data reduction procedures are discussed in \S 4.  The image atlas is presented in \S 5, and photometric measurements are given in \S 6. Section 7 contains a discussion of basic results, including comparisons with prior measurements, and a summary is given in \S 8.  A reference cosmology of $\Omega_{\rm \Lambda} = 0.72, \Omega_{\rm m} = 0.28$ and $H_{\rm 0} = 70$ km sec.$^{-1}$ Mpc$^{-1}$ is adopted, however we also take into account local non-cosmological effects by using the three-attractor model of \citet{2000ApJ...528..655M}.

\section{The GOALS Sample}

\nocite{1996ApJS..103...81C}

The IRAS RBGS contains a total of 179 LIRGs (log${\ts(L_{\rm IR}/L_\odot)}= 11.0 - 11.99$), and 22 ultra-luminous infrared galaxies (ULIRGs: log${\ts(L_{\rm IR}/L_\odot) \geq 12.0}$); these 201 objects comprise the GOALS sample \citep{2009PASP..121..559A}, a statistically complete flux-limited sample of infrared-luminous galaxies in the local universe.   In addition to the {\em Herschel} observations reported here, the GOALS objects have been the subject of an intense multi-wavelength observing campaign, including VLA 20 cm \citep{1990ApJS...73..359C,1996ApJS..103...81C}, millimeter wave spectral line observations of CO(1$\rightarrow$0) emission  \citep{1991ApJ...370..158S}, sub-millimeter imaging at 450 \mum~and 850 \mum~\citep{2000MNRAS.315..115D}, near-infrared images from 2MASS \citep{2006AJ....131.1163S}, optical and $K$-band imaging \citep{2004PhDT........18I}, as well as space-based imaging from the {\em Spitzer Space Telescope} (IRAC and MIPS, Mazzarella \etal 2017, in prep), {\em Hubble Space Telescope} (ACS, Evans \etal 2017, in prep.), {\em GALEX} \citep[NUV and FUV,][]{2010ApJ...715..572H}, and the {\em Chandra X-ray Observatory} \citep[ACIS,][2017 in prep]{2011A&A...529A.106I}.  Extensive spectroscopy data also exist on the GOALS sample, such as in the optical \citep{1995ApJS...98..129K} and with {\em Spitzer} IRS in the mid-infrared \citep{2013ApJS..206....1S}.  {\em Herschel} PACS spectroscopy was obtained in Cycles 1 and 2 targeting the [\cii]~157.7 \mum, \oi~63.2 \mum, and \oiii~88 \mum~emission lines and the OH 79 \mum~absorption feature for the entire sample, as well as the \nii 122 \mum~line in 122 GOALS galaxies \citep[][2017 in prep]{2013ApJ...774...68D,2014ApJ...788L..17D}.  In addition, {\em Herschel} SPIRE FTS spectroscopy were obtained to probe the CO spectral line energy distribution from $J=4\rightarrow3$ up to $J=13\rightarrow12$ for 93 of the GOALS objects \citep[][2017, submitted]{2014ApJ...787L..23L, 2015ApJ...802L..11L}, as well as the \nii 205 \mum~emission line for 122 objects \cite{2013ApJ...765L..13Z,2016ApJ...819...69Z}.

Out of the original list of 203 GOALS systems, two were omitted from our {\em Herschel} sample, making for a final tally of 201 objects.  IRAS F13097--1531 (NGC 5010) was part of the original RBGS sample of \citet{2003AJ....126.1607S}, however due to a revision in the redshift of the object it was much closer than thought.  This caused the resulting far-IR luminosity to drop significantly below the LIRG threshold of $10^{11} L_\odot$.  The other object we excluded from our sample is IRAS 05223+1908, which we believe is a young stellar object (YSO), due to the fact that its spectral energy distribution (SED) peaks in the submillimeter part of the spectrum.

Table \ref{tbl:BasicGoalsData} presents the basic GOALS information.  Column 1 is the index number of galaxies in the GOALS sample, and correspond to the same galaxies in Tables 2, 3, and 4.  Column 2 is the IRAS name of the galaxy, ordered by ascending RA.  Galaxies with the ``F" prefix originate from the {\em IRAS} Faint Source Catalog, and galaxies with no ``F" prefix are from the Point Source Catalog.  Column 3 is a list of common optical counterpart names.  Columns 4 and 5 are the {\em Spitzer} 8 \mum~centers of the system in J2000 from Mazzarella \etal (2017).  For galaxy systems with two or more components, the coordinate is taken to be the geometric midpoint between the component galaxies.  Column 6 gives the angular diameter distance to the galaxy in Mpc, from Mazzarella \etal (2017).  Column 7 is the map size used in the atlas, denoting the physical length of a side in each atlas image in kpc.  Column 8 is the systemic heliocentric redshift of the galaxy system, and Column 9 is the measured heliocentric radial velocity in km sec$^{-1}$, that corresponds to the redshift.  Both of these columns take into account cosmological as well as non-cosmological effects \citep[see][]{2000ApJ...528..655M}.  Finally Column 10 is the indicative 8--1000 \mum infrared luminosity in log$\ts(L_{\rm IR}/L_{\odot})$ of the entire system from \cite{2009PASP..121..559A}.  Similar to Columns 8 and 9, the $L_\mathrm{IR}$ values in Table 1 take into account the effect of the local attractors to $D_A$, than one would normally obtain from pure cosmological effect.

\def\tableBasicInfo{
\begin{deluxetable}{cllcrrclcc}
\tablecaption{Basic GOALS Data\label{tbl:BasicGoalsData}} 
\tabletypesize{\footnotesize} 
\singlespace
\tablenum{1}
\tablewidth{0pt}
\tablewidth{7.5in}

\tablehead{
\colhead{\#} & \colhead{IRAS Name} & \colhead{Optical Name} & \colhead{Right Ascension} & \colhead{Declination} & \colhead{$D_A$} & \colhead{Map Size} & \colhead{Redshift} & \colhead{Velocity} & \colhead{$L_{\rm IR}$} \\
\colhead{---} & \colhead{---} & \colhead{---} & \colhead{$\mathrm{HH:MM:SS}$} & \colhead{$\mathrm{DD:MM:SS}$} & \colhead{Mpc} & \colhead{kpc} & \colhead{---} & \colhead{km s$^{-1}$} & \colhead{log$\left(\frac{L}{L_\odot}\right)$} \\
\colhead{(1)} & \colhead{(2)} & \colhead{(3)} & \colhead{(4)} & \colhead{(5)} & \colhead{(6)} & \colhead{(7)} & \colhead{(8)} & \colhead{(9)} & \colhead{(10)}
}
\startdata
1 & F00073+2538 & NGC 23 & $00:09:53.36$ & $+25:55:27.7$ & $63.3$ & $100$ & $0.01523$ & $4566$ & $11.12$ \\
2 & F00085-1223 & NGC 34, Mrk 938 & $00:11:06.56$ & $-12:06:28.2$ & $81.5$ & $100$ & $0.01962$ & $5881$ & $11.49$ \\
3 & F00163-1039 & Arp 256, MCG-02-01-051/2 & $00:18:50.37$ & $-10:22:05.3$ & $111.4$ & $150$ & $0.02722$ & $8159$ & $11.48$ \\
4 & F00344-3349 & ESO 350-IG 038, Haro 11 & $00:36:52.49$ & $-33:33:17.2$ & $85.4$ & $100$ & $0.0206$ & $6175$ & $11.28$ \\
5 & F00402-2349 & NGC 232 & $00:42:49.32$ & $-23:33:04.3$ & $91.3$ & $150$ & $0.02217$ & $6647$ & $11.44$ \\
6 & F00506+7248 & MCG+12-02-001 & $00:54:03.88$ & $+73:05:05.9$ & $67.7$ & $100$ & $0.0157$ & $4706$ & $11.50$ \\
7 & F00548+4331 & NGC 317B & $00:57:39.72$ & $+43:47:47.7$ & $74.1$ & $100$ & $0.01811$ & $5429$ & $11.19$ \\
8 & F01053-1746 & IC 1623, Arp 236 & $01:07:47.54$ & $-17:30:25.6$ & $82.2$ & $100$ & $0.02007$ & $6016$ & $11.71$ \\
\enddata		
\small
\tablecomments{
The column descriptions are
(1) The row reference number.
(2) The IRAS name of the galaxy, ordered by ascending RA.  Galaxies with the ``F" prefix originate from the {\em IRAS} Faint Source Catalog, and galaxies with no ``F" prefix are from the Point Source Catalog.
(3) Common optical counterpart names to the galaxy systems.
(4) --- (5) are the {\em Spitzer} 8\mum~centers of the system in J2000 (see Mazzarella \etal 2017).  These {\em Herschel} atlas images are centered on these coordinates.  For widely separated systems, the coordinate is taken to be half way between the component galaxies.
(6) The angular diameter distance to the galaxy in Mpc, from Mazzarella \etal (2017).
(7) The map size used in the atlas, denoting the physical length of a side in each atlas image, in kpc.
(8) The non-relativistic redshifts as reported by NED, which also corresponds to the heliocentric velocity (Column 9), and includes both cosmological and non-cosmological effects.
(9) The measured heliocentric radial velocity corresponding to the redshift in Column 8 in km$/$sec from \citet{2009PASP..121..559A}, which includes both cosmological and non-cosmological effects.
(10) The indicative infrared luminosity, measured in log$(L/L_{\odot})$ of the entire system, from \cite{2009PASP..121..559A}.  These values take into account the effect of the local attractor to the distances than one would normally obtain from pure cosmological effects.
}
\tablenotetext{a}{These are very widely separated galaxy pairs that required two {\em Herschel} PACS observations.}
\end{deluxetable}
}
\clearpage
\begin{onecolumn}
\ifnum\Mode=0 
\placetable{tbl:BasicGoalsData}
\else
\ifnum\Mode=2 \onecolumn \fi 
\tableBasicInfo
\ifnum\Mode=2 \twocolumn \fi 
\fi 
\end{onecolumn}

\ifnum\Mode=0
\placefigure{fig:PacsSpireFilterCurves}
\else
\begin{figure*}[tb]
\figurenum{1}
\label{fig:PacsSpireFilterCurves}
\center
\vskip -0.2in
\includegraphics[scale=0.83,angle=0]{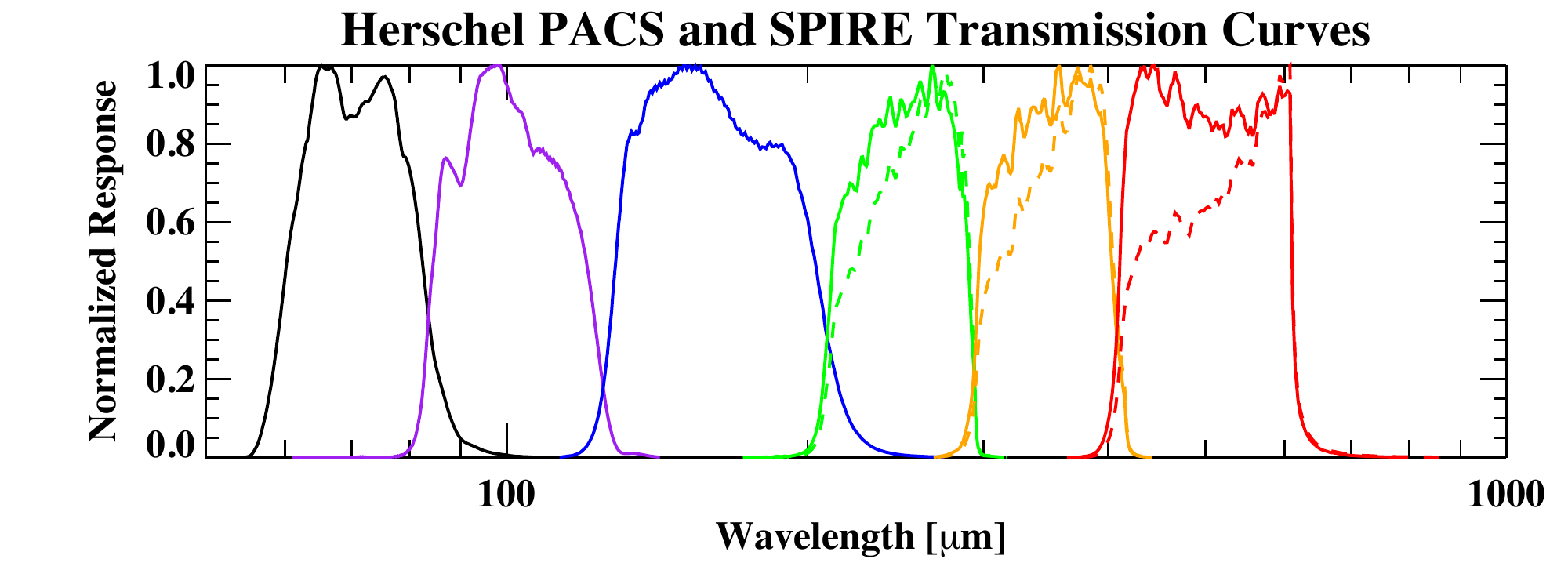}
\caption{The normalized filter transmission curves for our {\em Herschel} data.  From left to right are the PACS 70 \mum, 100 \mum, 160 \mum~channels, followed by the SPIRE 250 \mum, 350 \mum, and 500 \mum~channels.  For the SPIRE bands, the point source response is shown with a solid curve, while the extended source response is shown with a dashed curve.  Note the large difference in response for the SPIRE 500 \mum~transmission curve. }
\end{figure*}
\fi

\ifnum\Mode=0
\placefigure{fig:PacsSpireObservationExample}
\else
\begin{figure*}[tb]
\figurenum{2}
\label{fig:PacsSpireObservationExample}
\center
\vskip -1.0in
\includegraphics[scale=0.47,angle=0]{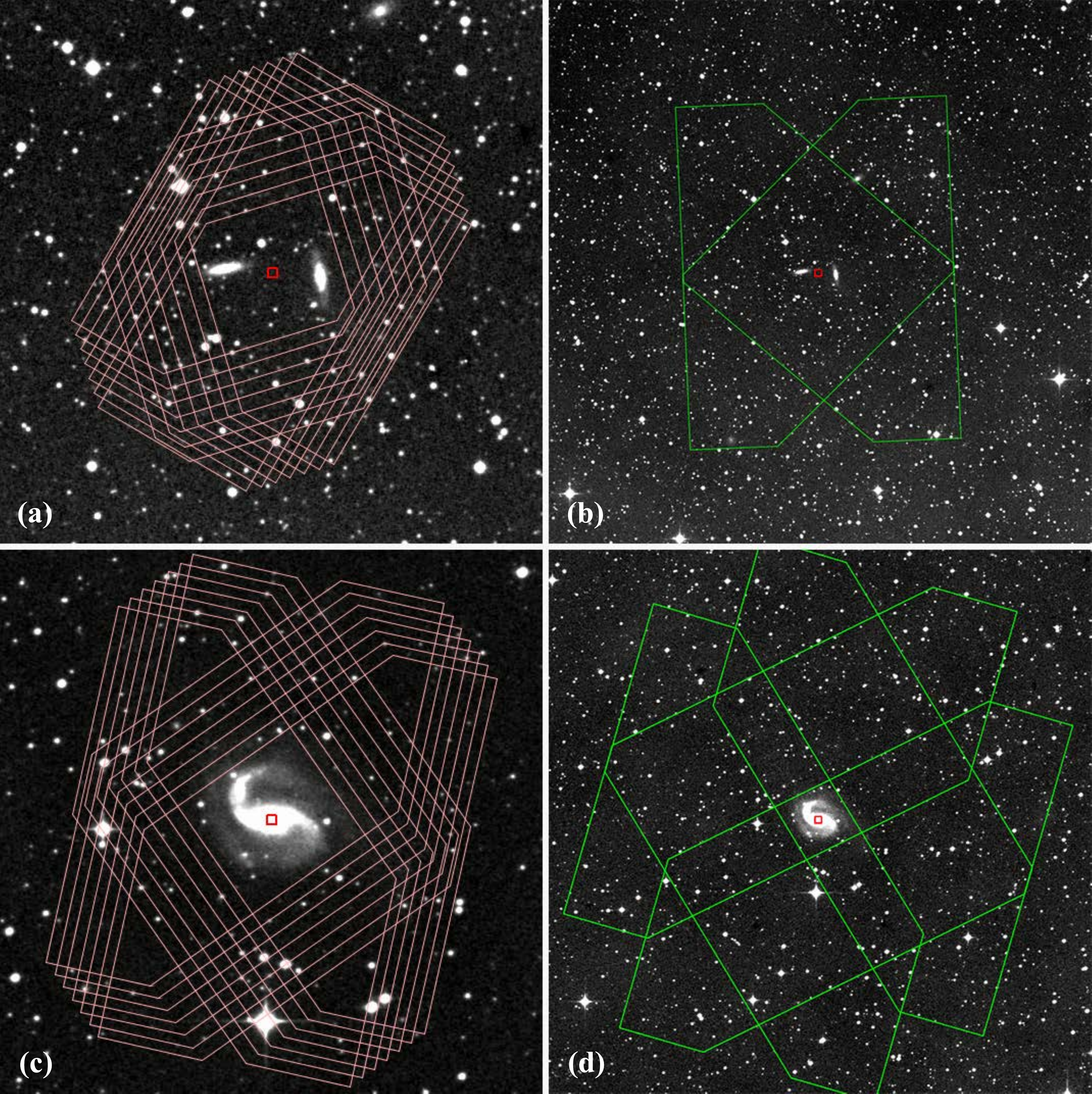}
\caption{The PACS and SPIRE observation footprints for two galaxies, IRAS F18145+2205 (CGCG 142-034) in the top row, and IRAS F20221-2458 (NGC 6907) on the bottom.  These figures were generated using HSPOT, the Herschel observation planning tool, while the background images used are from DSS.  The red box in each panel indicates the central coordinate for each observation.  The PACS observations are shown in panels (a) and (c), which show a 9\arcmin $\times$ 9\arcmin~field of view around the target coordinate.  Each scan leg in one direction is repeated several times (nominally 7 times) for maximal coverage of the source galaxy (or galaxies).  The SPIRE observations are shown in panels (b) and (d), and have a 25\arcmin $\times$ 25\arcmin~field of view, which is much larger than the PACS field of view.  Panel (b) shows a small map scan, while the bottom panel shows a large map scan.}
\end{figure*}
\fi

\section{{\em Herschel Space Observatory} Observations}
The {\em Herschel Space Observatory} \citep{2010A&A...518L...1P} imaging observations of the GOALS sample took place between the dates of March 2011 through June 2012, through our Cycle 1 open time observing program OT1\_dsanders\_1 (PI: D. Sanders, Program ID \#1279).  A total of 169 galaxy systems were observed by the Photoconductor Array Camera and Spectrometer \citep[PACS,][]{2010A&A...518L...2P} instrument in imaging mode under our proposal, with data from the remaining 32 galaxy systems from other guaranteed time (GT) or open time key programs (KPOT) obtained from the Herschel Science Archive (HSA).  In addition we observed 160 targets with the Spectral and Photometric Imaging Receiver \citep[SPIRE,][]{2010A&A...518L...3G}, with the remaining 41 targets from other GT and KPOT programs extracted from the HSA.  In total, 84.9 hours of observations were completed under our specific GOALS program, with 61.6 hours for PACS and 23.3 hours for SPIRE.

Broad-band imaging were obtained in the three PACS bands at 70 \mum, 100 \mum, and 160 \mum, and the three SPIRE bands at 250 \mum, 350 \mum, and 500 \mum.  The normalized filter transmission curves are shown in Figure \ref{fig:PacsSpireFilterCurves}.  Each SPIRE band has two curves associated with the filter, corresponding to the point source responsitivity (solid) and extended source responsitivity (dashed), which is important since some of the objects in our sample are extended even at SPIRE wavelengths (\ie the LIRG IRAS F03316--3618/NGC 1365).

Within the GOALS sample there are eight systems consisting of widely separated pairs where two separate PACS observations were needed, but only one SPIRE observation was made since its field of view was larger.  These galaxies are denoted in both Tables \ref{tbl:BasicGoalsData} and \ref{tbl:ObsLog}, giving a total of $201+8=209$ observation datasets.  We note for the galaxy system IRAS F07256+3355 which has three components, only two are visible in the PACS imagery, due to the smaller field of view of PACS.  The third component (NGC 2385) is far to the west and still within SPIRE's larger field of view.  Using the SPIRE fluxes as a rough proxy for infrared luminosity strength, NGC 2385 contributes very little to the overall infrared luminosity of the system.  IRAS F23488+1949 also has a third component (NGC 7769) to the NNW in the SPIRE images, but is outside of the PACS scan area.  However from the SPIRE fluxes NGC 7769 appears to have a moderate contribution to the system's infrared luminosity.  In sum we achieved a very high degree of coverage and completeness for each GOALS object with {\em Herschel}.

\subsection{Photoconductor Array Camera and Spectrometer (PACS) Observations}
The Photoconductor Array Camera and Spectrometer \citep[PACS,][]{2010A&A...518L...2P} is one of three far infrared instruments onboard the {\em Herschel Space Observatory} and covers a wavelength range between 60 -- 210 \mum.  In the photometer mode it can image two simultaneous wavelength bands centered at 160 \mum, and at either 70 \mum~or 100 \mum.  These three broad bands are referred to as the blue channel (60 -- 85 \mum), green channel (85 -- 130 \mum) and red channel (130 -- 210 \mum).  For any given observation, the blue camera observes at either 70 \mum~or 100 \mum, while the red camera only observes at 160 \mum.  A dichroic beam-splitter with a designed transition wavelength of 130 \mum~directs the incoming light into the blue and red cameras, and a filter in front of the blue camera selects either the blue or green band.

The detectors for both the blue and red cameras comprise a filled bolometer array of square pixels that instantaneously samples the entire beam from the telescope's optics.  The layout of the blue camera's focal plane consists of 4 $\times$ 2 subarrays, with 16 $\times$ 16 pixels in each subarray.  Similarly the red camera consists of 2 $\times$ 1 subarrays with 16 $\times$ 16 pixels each.  On the sky each bolometer pixel subtends an angle of 3\arcs2 $\times$ 3\arcs2 and 6\arcs4 $\times$ 6\arcs4 for the blue and red cameras respectively.  There exists gaps between each of the subarrays in both cameras, which must be filled in by on-sky mapping techniques (\ie, scan mapping).  Both the blue and red cameras were designed to image the same 3\arcm5 $\times$ 1\arcm75 field of view on the sky at any given instant.

In the photometer mode there are two astronomical observing templates (AOT) available, in addition to a PACS/SPIRE parallel observing mode.  For our {\em Herschel} GOALS program we used the scan map technique for all of our astronomical observing requests (AOR), which is ideal for mapping large areas of the sky and/or targets where extended flux may be present.  Our scan map observations involve slewing the telescope at constant speed along parallel lines separated by 15\arcsec~from each other, perpendicular to the scan direction.  Two example PACS observation footprints are shown in Figure \ref{fig:PacsSpireObservationExample} panels (a) and (c), overlaid on images from the Digital Sky Survey (DSS).  The area of maximum coverage is the inner region centered on the red box, where the requested observation is centered.  For the GOALS observations we chose to observe 7 scan legs in each scan and cross-scan using the 20\arcsec/sec scan speed, with scan leg lengths ranging between 3 -- 6\arcmin~depending on the size of the target.  At this scan speed the beam profiles for each wavelength have mean FWHM values of 5\arcs6, 6\arcs8, and 11\arcs3 for the 70 \mum, 100 \mum, and 160 \mum~channels respectively.

Before each PACS photometer observation is a 30 sec.~chopped calibration measurement between two internal calibration sources (the calibration block), followed by 5 sec.~of idle for telescope stability before the science observation is executed.  As the telescope is scanned across the sky during science observations, all of the bolometer pixels are read out at a frequency of 40 Hz, even during periods where the telescope was turning around for the next scan leg.  However due to satellite data-rate limitations, all PACS data are averaged over four frames effectively downsampling the data to 10 Hz.  The result is a data timeline of the flux seen by each detector pixel as a function of time (and by extension position on the sky) as the telescope is scanned over the target field.

In order to accurately reconstruct the image, two scan map AORs at orthogonal angles are required.  This is because as the telescope scans a field, the offsets of each bolometer subarray, and even each pixel, may be different from its neighbor resulting in stripes or gradients in the final reconstructed map.  However if the same field is scanned in two orthogonal directions, many of these map artifacts can be successfully removed, by virtue of multiple different bolometers sampling each patch of the sky.  Furthermore in order to maximally sample a given sky pixel by as many bolometer pixels as possible, we chose our scan angle to be 45$\degs$ and 135$\degs$ with respect to the detector array.  The orthogonal scans similarly help remove drifts in the bolometer timelines, which are time-dependent variations in the detector or subarray offsets, caused by for example cosmic ray hits and other instrument effects.  For our survey the typical PACS scan duration is about 200 sec., however larger maps with deeper coverage can be as long as $\sim$1900 sec.  Since two scans are needed for each target in the blue and green filter, there are two pairs of scan and cross-scan in the red channel, giving us better sensitivity.  Unfortunately due to unforeseen consequences, several galaxy components in IRAS F02071--1023 and IRAS F07256+3355 had sufficient coverage by only one of the scans, which resulted in more noise along the scan direction around the target.

Since by definition all of the objects in the GOALS sample have an {\em IRAS} 60 \mum~flux of at least 5.24 Jy, the galaxies or galaxy systems are bright enough such that only one repetition was needed for each PACS scan and cross scan.  With one pair of scan and cross-scan observations, we achieved a 1-$\sigma$ point source sensitivity of approximately 4 mJy in the central area, and approximately 8 mJy averaged over the entire map for both blue and green observations.  By combining all four red channel scans and cross-scans we achieved a 1-$\sigma$ point source sensitivity of about 6 mJy in the central area, and about 12 mJy averaged over the entire map.  On the other hand the extended flux sensitivities for one repetition (one scan and cross-scan pair) are 5.3 MJy sr$^{-1}$, 5.2 MJy sr$^{-1}$, and 1.7 MJy sr$^{-1}$ for the 70 \mum, 100 \mum, and 160 \mum~channels respectively.

\subsection{Spectral and Photometric Imaging Receiver (SPIRE) Observations}
The Spectral and Photometric Imaging Receiver \citep[SPIRE,][]{2010A&A...518L...3G} is a submillimeter camera on {\em Herschel} that operates between the 194-671 \mum~wavelength range.  In the imaging mode, it can simultaneously observe in three different broad bandpasses ($\lambda/\Delta\lambda\sim3$), centered at 250 \mum, 350 \mum, and 500 \mum.  Similar to PACS, SPIRE images a field by scan mapping, where the instrument field of view (4\arcmin $\times$ 8\arcmin) is scanned across the sky to maximize the spatial coverage.  The three detector arrays use hexagonal feedhorn-coupled bolometers, with 139, 88, and 43 bolometers for the PSW (250 \mum), PMW (350 \mum), and PLW (500 \mum) channels respectively.  The beam profiles for each wavelength have mean FWHM values of 18\arcs1, 25\arcs2, and 36\arcs6 for the 250 \mum, 350 \mum, and 500 \mum~photometer arrays, and mean ellipticities of 7\%, 12\%, and 9\% (the beam shape changes slightly as a function of off-axis angle).

There are three main observing modes available: point source photometry, field/jiggle mapping, and scan mapping.  For our observing program (dsanders\_OT1\_1) we chose the scan-map mode at a scan rate of 30\arcsec/sec., since it gave the best data quality and also larger field of view for the final map than the other two mapping modes.  Nominal scan angles of 42.4$\degs$ and 127.2$\degs$ with respect to the detector arrays were used to maximize sky coverage by as many detectors as possible, and to minimize the effect of individual bolometer drift during data processing.  Like PACS, two scans are needed for data redundancy as well as cross-linking, however the scan and cross-scan with SPIRE are observed within a single AOR.  Within our program, the vast majority of our targets were observed in the small map mode ($\sim$150 targets), while the rest were taken in the large map mode ($\sim$20 targets).  The typical scan durations are $\sim$170 sec.~for small maps ($\sim$5\arcmin $\times$ 5\arcmin~guaranteed map coverage area), and up to $\sim$2200 sec.~for large maps.  In Figure \ref{fig:PacsSpireObservationExample} panels (b) and (d) we show two example observations using SPIRE.  The top panel shows a small map mode observation, while the bottom panel shows a large map mode observation.  In both cases the SPIRE detector is scanned over the target coordinate (shown by the red box) from the top left to the bottom right, and then from the top right to the bottom left.

Due to the extremely sensitive design of the {\em Herschel} optics and SPIRE instrument, only one repetition was observed for every target in our observing program.  The SPIRE instrument has a confusion limit of 5.8, 6.3, and 6.8 mJy beam$^{-1}$ for the 250 \mum, 350 \mum, and 500 \mum~channels, which is defined as the standard deviation of the flux density in the limit of zero instrument noise \citep{2010A&A...518L...5N}. On the other hand the instrument noise is about 9, 7.5, and 10.8 mJy beam$^{-1}$ at 250 \mum, 350 \mum, and 500 \mum~for one repetition (scan and cross-scan) at the nominal scan speed of 30\arcsec/sec.  Since many of our targets have extended features, SPIRE's 1-$\sigma$ sensitivities to extended flux are at the 1.4 MJy sr$^{-1}$, 0.8 MJy sr$^{-1}$, and 0.5 MJy sr$^{-1}$ levels for 250 \mum, 350 \mum, and 500 \mum~for one repetition.  These flux levels are already dominated by confusion noise, and is more than enough to detect any cold dust components in our sample.

\subsection{Observing Log}
Table \ref{tbl:ObsLog} below lists the observing log for our data sample.  Column 1 is the galaxy reference number, and column 2 is the IRAS name of the galaxy, ordered by ascending RA.  Column 3 is the common optical counterpart names to the galaxy systems.  Columns 4 -- 7 are the observation IDs for PACS imaging.  Blue corresponds to a wavelength of 70 \mum, while green corresponds to 100 \mum.  Each blue and green observation pair simultaneously observes the red 160 \mum~channel.  Two orthogonal observations are made at each wavelength to reduce imaging artifacts.  We note that four galaxies in our sample do not have 100 \mum~observations available since they were from other programs that did not observe them: IRAS F02401-0013 (NGC 1068), IRAS F09320+6134 (UGC 05101), IRAS F15327+2340 (Arp 220), and IRAS F21453-3511 (NGC 7130).  Column 8 is the PACS observation duration for each scan and cross-scan, unless otherwise noted.  We note these are {\em not} exposure times, but instead the amount of time for each scan and cross-scan.  Columns 9 -- 10 are the observation dates (in YYYY-MM-DD) for each pair of PACS scan and cross-scan, unless otherwise noted, while column 11 is the Program ID of the PACS program from which the data were obtained.  We list the PID corresponding to each number in Table \ref{tbl:ObsLog}'s caption.  The bulk of the data ($\sim$80\%) are from OT1$\_$sanders$\_$1, with most of the remaining data from KPGT$\_$esturm$\_$1 and KPOT$\_$pvanderw$\_$1.  Column 12 is the SPIRE observation ID which includes all three 250 \mum, 350 \mum, and 500 \mum~observations.  The scans and cross-scans for each target is combined into one observation.  Column 13 is the SPIRE observation duration, which is similar to the PACS duration.  Column 14 is the SPIRE observation date, and column 15 is the PID of the SPIRE program from which the data were obtained, similar to the PACS PID column.

\def\tableObsLog{
\begin{deluxetable}{cllccccccccccccc}
\tabletypesize{\footnotesize} 
\setlength{\tabcolsep}{0.04in} 
\rotate
\singlespace
\tablenum{2}
\tablewidth{0pt}
\tablecaption{Herschel Observation Log}\label{tbl:ObsLog}

\tablehead{
\multicolumn{1}{c}{} &  
\multicolumn{1}{c}{} &
\multicolumn{1}{c}{} &
\multicolumn{8}{c}{PACS}   &
\multicolumn{1}{c}{} &  
\multicolumn{4}{c}{SPIRE} \\
 
\cline{4-11} \cline{13-16} \\
\multicolumn{1}{c}{\#} &  
\multicolumn{1}{c}{IRAS Name} &
\multicolumn{1}{c}{Optical Name} &
\colhead{Blue 1} &
\colhead{Blue 2} &
\colhead{Green 1} &
\colhead{Green 2} &
\colhead{Duration} &
\colhead{Blue} & 
\colhead{Green} & 
\colhead{PACS} & 
\colhead{} & 
\colhead{Obs. ID} &
\colhead{Duration} &
\colhead{Obs. Date} &
\colhead{SPIRE} \\ 

\colhead{} &
\colhead{} &
\colhead{} &
\colhead{Obs. ID} &
\colhead{Obs. ID} &
\colhead{Obs. ID} &
\colhead{Obs. ID} &
\colhead{(sec.)} &
\colhead{Obs. Date} & 
\colhead{Obs. Date} & 
\colhead{PID\tablenotemark{1}} & 
\colhead{} & 
\colhead{} &
\colhead{(sec.)} &
\colhead{} &
\colhead{PID\tablenotemark{1}} \\ 

\colhead{(1)} &     
\colhead{(2)} &     
\colhead{(3)} &     
\colhead{(4)} &     
\colhead{(5)} &     
\colhead{(6)} &     
\colhead{(7)} &     
\colhead{(8)} &     
\colhead{(9)} &     
\colhead{(10)} &    
\colhead{(11)} &    
\colhead{} &          
\colhead{(12)} &    
\colhead{(13)} &    
\colhead{(14)} &    
\colhead{(15)}       
}
\startdata
1 & F00073+2538 & NGC 23 & 1342225471 & 1342225472 & 1342225469 & 1342225470 & 198 & 2011-07-24 & 2011-07-24 & 1 & ~ & 1342234681 & 169 & 2011-12-18 & 1 \\
2 & F00085-1223 & NGC 34, Mrk 938 & 1342212463 & 1342212464 & 1342212465 & 1342212466 & 276 & 2011-01-10 & 2011-01-10 & 3 & ~ & 1342199384 & 445 & 2010-06-29 & 3 \\
3 & F00163-1039 & Arp 256, MCG-02-01-051/2 & 1342212700 & 1342212701 & 1342212702 & 1342212703 & 485 & 2011-01-15 & 2011-01-15 & 2 & ~ & 1342234694 & 169 & 2011-12-18 & 1 \\
4 & F00344-3349 & ESO 350-IG 038, Haro 11 & 1342210636 & 1342210637 & 1342197713 & 1342197714 & 65\tablenotemark{b} & 2010-12-01 & 2010-06-04 & 10 & ~ & 1342199386 & 307 & 2010-06-29 & 10 \\
5 & F00402-2349 & NGC 232 & 1342238073 & 1342238074 & 1342237438 & 1342237439 & 240 & 2012-01-21 & 2012-01-13 & 1 & ~ & 1342234699 & 1217 & 2011-12-18 & 1 \\
6 & F00506+7248 & MCG+12-02-001 & 1342237174 & 1342237175 & 1342237172 & 1342237173 & 198 & 2012-01-11 & 2012-01-11 & 1 & ~ & 1342199365 & 169 & 2010-06-29 & 4 \\
7 & F00548+4331 & NGC 317B & 1342213944 & 1342213945 & 1342213946 & 1342213947 & 347 & 2011-02-08 & 2011-02-08 & 2 & ~ & 1342238255 & 169 & 2012-01-27 & 1 \\
8 & F01053-1746 & IC 1623, Arp 236 & 1342212754 & 1342212755 & 1342212846 & 1342212847 & 65 & 2011-01-16 & 2011-01-18 & 2 & ~ & 1342199388 & 169 & 2010-06-29 & 2 \\
\enddata
\end{deluxetable}
}

\clearpage
\ifnum\Mode=0 
\vskip 0.3in
\vskip 0.3in
\placetable{tbl:no2}
\else
\ifnum\Mode=2 \onecolumn \fi 
\tableObsLog
\ifnum\Mode=2 \twocolumn \fi 
\fi 

\section{Data Processing and Reduction}
The data processing for our {\em Herschel} data was performed using the Herschel Interactive Processing Environment \citep[HIPE,][]{2010ASPC..434..139O} version 14 software tool, which provides the means to download, reduce, and analyze our data.  All of our data reduction routines are derived from the standard pipeline scripts found within HIPE, where the programming language of choice is Jython (a Java implementation of the popular Python language).  In addition to handling the data processing, HIPE also downloads and maintains all of the instrument calibration files needed for the data processing.

\subsection{PACS Data Reduction}
\subsubsection{Choosing a PACS Map Maker}
Due to the bolometer and scanning nature of the PACS instrument, it was important to determine the best map-making software to translate the time ordered data (TOD) into an image.  The PACS bolometers (indeed all bolometers) produce noise that increases as one approaches lower temporal frequencies, commonly referred to as $1/f$ noise, that must be removed by the map-maker.  If this noise is left uncorrected in the time ordered data, the result would be severe striping or even gradients across the image.  In addition the map making software must also remove the bolometers' common mode drift (which is a changing offset as function of time) from the TOD, termed {\em pre-processing}, as well as cosmic ray hits and individual bolometer drift.  The PACS team released a Map-making Tool Analysis and Benchmarking report\footnote{\tt{\url{http://herschel.esac.esa.int/twiki/pub/Public/
PacsCalibrationWeb/pacs\_mapmaking\_report\_ex\_sum\_v3.pdf}}} in November 2013 with an update in March 2014 that characterized in detail the six different map making packages available to reduce PACS data.  We summarize the information presented in this report below to decide upon the best map making software to use, since it was important that all of the {\em Herschel} data on our sample were processed uniformly.

The PACS team tested the performance of six different publicly available map-making packages: MADMap \citep[Microwave Anisotropy Dataset mapper,][]{2010ApJS..187..212C}, SANEPIC \citep[Signal And Noise Estimation Procedure Including Correlation,][]{2008ApJ...681..708P}, Scanamorphos \citep{2013PASP..125.1126R}, JScanam (Jython Scanamorphos\footnote{This is the HIPE/PACS implementation of the Scanamorphos algorithm however they both differ in many assumptions, hence why they were tested separately.}), Tamasis \citep[Tools for Advanced Map-making, Analysis and SImulations of Submillimeter surveys,][]{2011A&A...527A.102B}, and Unimap \citep{2015MNRAS.447.1471P} (see \S2 and \S4 of the Map-making Tool and Analysis Benchmarking report for a description of each code).  We did not consider the PACS high pass filter (HPF) reduction software, since HPF maps are background-subtracted and will miss a significant amount of extended emission outside approximately one beam area.  To evaluate each of the packages, a combination of simulated and real data from PACS were used.  Except in a few cases, most of our fluxes are within the Benchmarking report's ``bright flux regime" of $0.3$ -- $50$ Jy, while the ``faint flux regime" is defined to be $0.001$ -- $0.1$ Jy (see Figure \ref{fig:FluxHistogram}).  Below we summarize the five tests performed on each map maker from the benchmarking report:

1) A power spectrum analysis which tests the map maker's ability to remove noise while preserving extended fluxes over large angular sizes on the map.  This tests each code's performance in removing the $1/f$ noise from the PACS data, and consequently how well gradients and stripes are removed from the maps.

2) A difference matrix is computed for each map maker's output, which evaluates differences in fluxes for individual sky pixels over the entire image.  $(S-S_\mathrm{true})$ is computed for each pixel and plotted against $S_\mathrm{true}$, and the resulting scatter, offset, and slope is evaluated.

3) Each map maker's performance in point source photometry is compared to fluxes measured from the HPF maps for both bright ($0.3$ -- $50$ Jy) and faint ($0.001$ -- $0.1$ Jy) cases.  Since the HPF maps produced by HIPE are designed specifically for the case of point sources, they provide the most accurate reference point source fluxes.

4) Extended source photometry tests each map maker's ability to recover extended flux over large areas of the map.  To assess this, each code's output is compared to IRAS data on M31 from the Improved Reprocessing of the IRAS Survey \citep[IRIS,][]{2005ApJS..157..302M}.

5) The noise characteristics each map maker introduces into the final map are evaluated.  This includes statistical tests on the pixel-to-pixel variance as well as the shape of the overall distribution of fluxes in each map pixel.  The noise patterns are also evaluated with regard to how isotropic the noise appears in the maps.

Considering the results of these extensive tests, it was difficult to select the best map maker for our PACS data.  We rejected the High Pass Filter method outright since many of our galaxies are easily resolved at the PACS wavelengths, and would therefore have a significant amount of extended flux missed by the HPF pipeline.  We decided against using SANEPIC since it significantly overestimated the true flux for both bright and faint point source photometry.   We also ruled out using Tamasis since it has a tendency to introduce more pronounced noise along the scan directions.  This left us with four remaining choices: JScanam, MADMap, Scanamorphos, and Unimap.  We finally decided on using JScanam to reduce all of our PACS data, as it gave the best balance between photometric accuracy and map quality.  Specifically, it reproduced a power spectrum closest to the original, it had the flattest $(S-S_\mathrm{true})$ vs. $S_\mathrm{true}$ plot, and it yielded the most accurate photometry for both point and extended sources in both channels.

For a small fraction of our maps where JScanam could not remove all of the image artifacts (usually gradients due to non-optimal baseline subtraction), we used Unimap to process the data, since it performed just as well as JScanam.  Unlike JScanam, Unimap approaches map making differently, using the Generalized Least Square (GLS) approach, which is also known as the Maximum Likelihood method if the noise has a Gaussian distribution.  For a very few cases where even Unimap did not produce optimal results, we resorted to using MADMap. This map maker requires that the noise properties of the detectors are determined {\em a-priori}, from which a noise filter can be generated to filter out the $1/f$ noise.  Finally, despite that not all of the PACS maps were generated using the same map maker, we note that the resulting photometry from all three map makers are remarkably consistent as shown in the Benchmarking report (and addendum) from the PACS team, hence giving one the freedom to use the map maker that produces the best image quality.

\subsubsection{PACS Map Making With JScanam}
All of our {\em Herschel}-GOALS PACS data were reduced in HIPE 14 using the latest available PACS calibration version 72\_0 released in December 2015.  In order to alleviate the processing time for all 211 objects, we started our data processing from the Level 1 products downloaded from the HSA.  These Level 1 data products have the advantage of an improved reconstruction of the actual {\em Herschel} spacecraft pointing, which reduces distortions on the PSF due to jitter effects.  Compared to previous maps from our data processing, the new maps have slight shifts of up to $\sim$1\arcs5, and slightly smaller PSFs in unresolved GOALS objects.

Since each PACS scan and cross-scan are separate observations, JScanam requires two observations each for the blue and green data.  On the other hand the red channel data are observed simultaneously regardless if the blue or green filter is used, so we have four observations in the red channel.  Processing for both the blue and red cameras are identical, with the red data requiring a further step of combining the two pairs of scan and cross-scan data.  Below we describe the key steps in the data reduction process.

After loading each scan and cross-scan observation context from HSA into HIPE, the first step was to execute the task {\tt photAssignRaDec} to assign the RA and declination coordinates to each pixel in each frame which allows JScanam to run faster.  The next step was to remove the unnecessary frames taken during each turnaround in the scan or cross-scan using the {\tt scanamorphosRemoveTurnarounds} task.  We opted to use the default speed limit which is $\pm50\%$ of our nominal scan speed (20\arcsec/sec.), so any frames taken at scan speeds below 10\arcsec/sec. or above 30\arcsec/sec. were removed.  After turnaround removal the {\tt scanamorphosMaskLongTermGlitches} task in JScanam goes through the detector timelines and masks any long term glitches.

At this point we have a detector timeline of flux detected by the bolometers as a function of time with the turnarounds and long term glitches removed.  Using the {\tt scanamorphosScanlegBaselineFitPerPixel} task, our next step is to subtract a linearly fit baseline from each bolometer pixel of every scan leg, with the intention of creating a ``naive" map for source masking purposes.  This is done iteratively where the most important parameter is the {\tt nSigma} variable, which controls the threshold limit for source removal.  For our data any points above {\tt nSigma}$=$$2$ times the standard deviation of the unmasked data are considered real sources, until the iteration converges.

The next step is to join the scan and cross-scan data together for a higher signal-to-noise map to create the source mask.  In the {\tt scanamorphosCreateSourceMask} task we set a {\tt nSgima}$=$$4$, so that any emission above 4 standard deviations is masked out.  At this point it is not necessary to mask out all of the faint extended emission, only the brightest regions.  After the source mask is determined, they are applied to the individual scan and cross-scan timelines and the real processing begins.

With the {\tt galactic} option set to ``true" in {\tt scana-
morphosBaselineSubtraction}, we only want to remove an offset in the time ordered data over all the scan legs, and subtract it from all the frames.  This is done by calculating a median offset over only the unmasked part of the data which importantly does not include any bright emission, and subtracting it from each pixel's timeline.  This is so that extended flux is treated correctly when subtracting the baseline (due to the telescope's own infrared emission) from the signal timelines, even in cases where the emission is not concentrated in a small region.  We emphasize this does not imply the subtraction of the Galactic foreground emission from our maps.

Once the baseline is removed we need to identify and mask the signal drifts produced by the calibration block observation.  In previous versions of our reduction, these drifts have produced very noticeable gradients in our final maps.  To do this the task {\tt scanamorphosBaselinePreprocessing} assumes that the scan and cross-scan are orthogonal to each other, which would result in gradients in different directions.  The drift removal is also based on the assumption that the drift power increases with the length of the considered time (1/$f$ noise).  For this reason the first iteration removes the drift component over the longest time scale which corresponds to the entire scan (or cross-scan).  After that drifts are removed over four scan legs, and finally over one scan leg, with the remaining drift in each successive iteration becoming weaker.  In order to actually calculate the drift in each iteration, a single scan (or scan legs) is back projected over itself in the orthogonal direction, which transforms the generally increasing or decreasing signal drift into oscillatory drifts that cancel out on large time scales.  The orthogonal back projected timeline is then subtracted from the scan timeline, and the difference which represents the drift is fitted by a line.

At this point the scan and cross-scan data have been cleaned enough to be combined.  Since signal drifts were only eliminated over timescales down to one scan leg, the next step is to remove them from over time scales shorter than one scan leg.  These drifts are due to for example cosmic ray hits on the PACS instruments, which produce different effects on the time ordered data depending on which part is hit.  If an individual bolometer or bolometer wall is hit, it only affects those bolometer(s).  However if a cosmic ray hits the readout electronics, it introduces a strong positive or negative signal for all of the bolometers read by the electronics, which can be anything from a single bolometer to an entire detector group.  These jumps typically last a few tens of seconds before settling to the previous level again, and would result in stripes across the final map if not properly removed.

To remove these individual drifts, we use the task {\tt scanamorphosIndividualDrifts} to first measure the scan speed and calculate the size of a map pixel that can hold six subsequent samples of a detector pixel crossing it.  We use a threshold of {\tt nSigma}$=$$5$ which is large enough to include the strongest drifts but still masking out the real source.  Then the average flux value and standard deviation from the detector pixels crossing that map pixel is calculated, along with the number of detector pixels falling into that map pixel.  Using the threshold noise value (from the calibration files), we eliminate any individual detector fluxes for that map pixel that has a standard deviation greater than the noise threshold.  The missing values are then linearly interpolated, and the individual drift is subtracted from the detector timeline.

After all of the individual drifts are corrected, the time ordered data are saved and we project the timelines from both the scans and cross-scans into our final map using the {\tt photProject} task.  We use a pixel scale of 1\arcs6 pixel$^{-1}$ for the 70 \mum~and 100 \mum~maps, and a pixel scale of 3\arcs2 pixel$^{-1}$ for the 160 \mum~maps.  By default the {\tt photProject} task assumes in projection an active pixel size of 640 \mum, however if we `drizzle' the projection we can assume smaller PACS pixels.  This allows us to reduce the noise correlation between neighboring map pixels and also sharpens the PSF.  We used a {\tt pixfrac} of $0.1$, which controls the ratio between the input detector pixel size and the map pixel size.  At this point the 70 \mum~and 100 \mum~maps are finished.  For the 160 \mum~data, both pairs of scan and cross-scan are identically processed separately, and then combined in the end using {\tt photProject} again.

\subsection{SPIRE Data Reduction}
\subsubsection{Choosing a SPIRE Map Maker}
Similar to the PACS instrument, the SPIRE detectors exhibit certain effects that are characteristic to bolometers.  Namely, they introduce an increasing amount of noise as the length of the considered time increases ($1/f$ noise), as well as constant and changing offsets (drifts) which could result in stripes or gradients in the final image.  Therefore any map maker for SPIRE must be able to remove these instrumental effects, while preserving flux (point source and extended) and creating distortion-free maps.  The SPIRE team released a Map Making Test Report\footnote{http://arxiv.org/pdf/1401.2109v1.pdf} in January 2014 that benchmarked in depth seven different map making codes, several of which were also present in the PACS Map Making report.  The map makers that participated in the benchmarking were the Naive Mapper, Destriper in two flavors (P0 and P1), Scanamorphos, SANEPIC, and Unimap.  The two flavors of the Destriper differ in the polynomial order used to subtract the baseline, where P0 corresponds to a polynomial order of 0 (\ie the mean) and P1 corresponds to an order of 1.  Two additional super-resolution map makers were also tested, however we did not consider them for processing our SPIRE data.  For a summary of each map maker we refer the reader to the SPIRE Map Making Test Report.

For the Map Making Test Report, the authors tested these five map makers based on a variety of benchmarks that are very similar to the PACS Map-Making Tool Analysis and Benchmarking report.  A combination of real and simulated SPIRE data were used, covering the full variety of science cases such as faint vs.~bright sources, extended vs.~point sources, and complex vs.~empty fields.  The simulated SPIRE data have the advantage of comparing each of the map makers' outputs to the ``truth" image, allowing for an unbiased comparison between all of the map making codes.  These simulated observations were synthesized from two different layers: a truth layer based on a real or artificial source, and a noise layer from real SPIRE observations so that both instrumental and confusion noise is accurately represented.  Below we summarize the four metrics and performance results for the five possible map makers:

1) Using simulated data, the deviation of each map maker's output is compared to the original synthetic data.  To quantify the deviation from truth, a scatter plot of $(S-S_\mathrm{true})$ is plotted against $S_\mathrm{true}$, and the resulting slopes, relative deviations, and absolute deviations are compared.

2) The 2D power spectrum of each map makers' output is compared to the ``truth" image.  The goal here is to quantify how well $1/f$ noise is removed from the maps while leaving real fluxes (point and extended) intact, as well as how high spatial frequency (small spatial scale) fluxes are treated.

3) Using the simulated data, point source photometry from each of the map makers were compared to the ``truth" images.  This tests how well point source fluxes are recovered by each map maker in both the bright ($S\approx300$ mJy) and faint ($S\approx30$ mJy) regimes.

4) Finally, extended source photometry was tested between all the map makers using the synthetic data.  A simulated exponential disk with an $e$-folding radius of 90\arcsec was used, and fluxes were measured using aperture photometry.

Using the results from these tests, we concluded that the best map maker to use was the Destriper P0 mapper.  It performed remarkably well among the other map makers, especially in cases where complex extended emission is present.  Although the Map Making report warned about its inability to properly remove the ``cooler burp" effect, the most recent version of Destriper P0 in HIPE 14 was updated to include proper treatment of this instrument effect.  On the other hand Destriper P1 compared unfavorably, especially in introducing artificial gradients in many cases.  The Naive Mapper was also ruled out due to it frequently over-subtracting the background where extended emission is present.  The map maker SANEPIC showed significant deviations from the ``truth" map, because the code makes some incorrect assumptions about the data.  Finally, although Scanamorphos can handle faint pixels very well, it showed significant deviations in the bright pixel case ($S>0.2$ Jy).  This is important since many galaxies in our sample are nearby and thus quite bright.

In HIPE 14, we used a more advanced version of the Destriper code called the ``SPIRE 2-Pass Pipeline" that was released by the SPIRE instrument team.  The basic pipeline processing steps and settings follow exactly that of the Destriper P0 (or P1 if the user so chooses) map maker, with the added benefit of producing exceptionally clean maps to be used in the final {\em Herschel} Science Archive.  Specifically, the 2-Pass Pipeline mitigates residual faint tails and glitches in the timeline, which if not removed can produce ringing effects.  The primary aim of this pipeline is to produce maps with better detections of outliers in the TOD such as glitches, glitch tails, and signal jumps, and remove any Fourier ringing that would result from failed outlier detections.  As an overview, the first pass runs a stripped down version of the pipeline using only the bare minimum tasks that excludes any Fourier analysis.  This includes running the Second Level Deglitching task to produce a mask over the glitches, which is then applied back to the Level 0.5 products\footnote{The Level 0.5 products are the output after running the raw satellite telemetry through the engineering pipeline.}.  Then a second pass of the pipeline is executed identical to the original Destriper map maker.

\subsubsection{SPIRE Map Making With 2-Pass Pipeline and Destriper P0}
Our final SPIRE maps were reduced in HIPE 14 using the latest calibration version {\tt SPIRECAL\_14\_2} released in December 2015.  Below we summarize the key data reduction steps, however a more detailed description on the photometer pipeline can be found within \citet{2010SPIE.7731E..36D}.

Our data processing begins with the Level 0 data products downloaded from HSA, which are the raw data formatted from satellite telemetry containing the readout in ADU from each SPIRE bolometer.  After an observation is loaded into HIPE, the first step is to execute the Common Engineering Conversion and format it into Level 0.5 products.  These products are the uncalibrated and uncorrected timelines measured in Volts, and contain all of the necessary information to build science-grade maps.

The first step in processing our data from Level 0.5 to Level 1 is to join all the scan legs and turnarounds together.  The turnaround occurs when the spacecraft turns around after a scan leg to begin another scan.  We opted to use the turnaround data to include as much data within our maps as possible.  Next the pipeline produces the pointing information for the observation, based on the positions of the SPIRE Beam Steering Mechanism as well as the offset between SPIRE and the spacecraft itself (referred to as the Herschel Pointing Product).  This results in the SPIRE Pointing Product which is used later on in the pipeline.  After calculating the pointing information, the pipeline corrects for any electrical crosstalk between the thermistor-bolometer channels.  The thermistors measure the temperature of the array bath as a function of time so that later we may accurately subtract the instrument thermal contribution, or temperature drift from the data timelines.

The next step is the signal jump detector, which detects and removes jumps in the thermistor timelines that would otherwise cause an incorrect temperature drift correction.  To do this, the module subtracts baselines and smoothed medians from the thermistor timelines to identify any jumps.  After deglitching the thermistor timelines, we must deglitch any cosmic ray hits on the bolometers themselves.  This is an important step since any glitches that are not removed would manifest itself as image artifacts on the final maps.  The pipeline does this in two steps, where the first step is to remove glitches that occur simultaneously in groups of connected bolometer detectors.  This can occur when a cosmic ray hits the substrate of an entire photometer array, and can leave an imprint of the array on the final map.  The second step is to run the wavelet deglitcher on the timeline data, which uses a complex algorithm to remove glitches in Fourier space.

After deglitching the detector timelines, a low pass filter response correction is applied to the TOD.  This is to take into account the delay in the electronics with respect to the telescope position along a scan, in order to ensure a match between the astrometric timeline from the telescope, and the detector timeline from the instrument.  At this point we can apply the flux conversion to the detector timelines, changing the units from Volts to Jy beam$^{-1}$.  The next step involves corrections to the timelines due to temperature drifts, which are caused by variations of the detector array bath temperatures.  First, with the {\tt coolerBurpCorrection} flag set to {\tt true},  the pipeline flags data that were affected by the ``cooler burp" effect.  Observations taken during this effect, usually in the first $\sim$8 hours of SPIRE observations, can create unusual temperature drifts.  The temperature drift correction step then removes low frequency noise by subtracting a correction timeline for each detector using data and calibration information.  The ``cooler burp" is also removed at this stage by applying additional multiplicative factors to the correction timeline.

Next we apply a bolometer time response correction which corrects any remaining low-level slow response from the bolometers.  This is done by multiplying the timelines in Fourier space by an appropriate transfer function obtained from a calibration file containing the detector time constants.  After this step we attach the RA and declination to the data timelines by using the SPIRE Pointing Product generated earlier.  Since many of our objects are extended in nature, we must apply an additional extended emission gain correction for individual SPIRE bolometers.  This is because the pipeline so far assumes uniform beams across the array, whereas in reality there exists small variations among different bolometers due to their positions on the array.

We then use the Destriper to remove striping from the final maps.  Since the dominant fluxes seen by SPIRE are from the telescope itself, the science signal is very small in comparison.  Therefore to isolate the science signal we must subtract out thermal contributions from the telescope.  However even after doing this, there are still large differences in residual offsets between different bolometers due to variations in the thermal and electronic aspects of the system, resulting in striping.  This is where the Destriper P0 comes in, which effectively takes as input SPIRE Level 1 context, and outputs destriped Level 1 timelines.  To do this we first subtract a median baseline as an initial guess, then we use a polynomial order of 0 to iteratively update the offsets in the TOD for each detector until an optimal solution is found.  This algorithm effectively normalizes the map background to zero, however we do include the true background using data from the {\em Planck} High Frequency Instrument (HFI) for the PMW and PLW arrays (see \S6.2.1).  After destriping we run the optional second level deglitching in order to remove any residual glitches that may still remain.

At this point the data have been processed to Level 1, and in the case of the first pass, only tasks that don't involve any manipulation in Fourier space were omitted.  The resulting second level deglitching mask from the first pass of this pipeline is applied to the Level 0.5 data, and the entire process is repeated in a second pass, this time including operations in Fourier space.

The final step in our SPIRE data reduction is to project the drift-corrected, deglitched, and destriped timelines into our Level 2 science grade map.  To do this we use a Naive Mapper, which simply projects the full power seen by a bolometer onto the nearest map pixel.  The final map pixel scales used were 6\arcsec, 10\arcsec, and 14\arcsec~for the PSW, PMW, and PLW arrays respectively.  For each instant of time on each bolometer's timeline, the measured flux is added to the total signal map and a value of 1 is added to the coverage map.  Once this is done for all bolometer timelines, the total signal map is divided by the coverage map to obtain the flux density map.

Although the 2-pass pipeline does an excellent job of removing all SPIRE image artifacts, approximately twenty of the maps still exhibited stripes and residual glitches in the final map.  These maps were reprocessed by first using the SPIRE bolometer finder tool to identify the misbehaving bolometer, and then masking the affected portions in that bolometer's Level 1 timeline.  The data were then rerun through the Naive Mapper to produce a clean and deglitched Level 2 science grade map.

\section{The {\em Herschel}-GOALS Image Atlas}
In the following pages in Figure \ref{fig:atlas} we present the entire {\em Herschel} atlas of the GOALS sample, ordered by ascending RA.  The archived\footnote{\url{http://irsa.ipac.caltech.edu/data/Herschel/GOALS}} {\em Herschel} GOALS maps are in standard {\tt *.fits} format with image units of Jy pixel$^{-1}$. Each page consists of six panels for the 70 \mum, 100 \mum, 160 \mum, 250 \mum, 350 \mum, and 500 \mum~channel maps.  

The IRAS name of each galaxy or galaxy system is shown at the top, along with their common names from optical catalogs.  Each of the six panels are matched and have the exact same map center as well as field of view.  The center coordinates of the {\em Herschel} atlas images are listed in Table \ref{tbl:BasicGoalsData}.  For galaxy systems with multiple components, the center coordinate is chosen to be roughly equidistant from all components.  The field of view for each panel is shown on the bottom left of the 70 \mum~panel, and represents the physical length of one side of each panel.  A scale bar also indicates the physical length of 10 kpc at the distance of the galaxy (derived from the angular diameter distance in Table \ref{tbl:BasicGoalsData}), along with the equivalent angular distance.  The circle on the bottom right of each panel represents the beam size at that wavelength.  Finally the right ascension and declination coordinates are indicated in J2000 sexagesimal as well as decimal format.  The sexagesimal RA coordinates have the hour portion truncated for all but the center tick mark, to keep the tick name sizes manageable.

Since many objects appear as point sources at some or all of the {\em Herschel} wavelengths, the morphologies of these galaxies will be dominated by the PSF at that wavelength.  In the case of PACS, the PSF is characterized by a narrow circular core elongated in the spacecraft $z$-direction, at 70 \mum~and 100 \mum.  In addition there is a tri-lobe pattern at the several percent level at all three wavelengths, however it is strongest at 70 \mum.  Finally, there are knotty structured diffraction rings at the sub-percent level, again most apparent at 70 \mum~and 100 \mum.  In the case of SPIRE, the PSF appears mostly circular, however for the brightest objects, airy rings are also visible.

In order to show as much detail in these maps, we used an inverse hyperbolic sine (asinh) stretch function to maximize the dynamic range of visible structures.  Also to keep all the PACS images uniform, the background for each image was adjusted such that the background is very close to zero.  The format in our {\em Herschel} atlas matches that of companion image atlases from {\em Hubble Space Telescope}-ACS (Evans \etal 2017) and {\em Spitzer}-IRAC/MIPS (Mazzarella \etal 2017), allowing one to study the morphological properties of these galaxies from 0.4 \mum~to 500 \mum.

\def\figcapAtlasOne{
\footnotesize 
The {\em Herschel} GOALS atlas, displaying imagery of local LIRGs and ULIRGs in the three PACS bands and three SPIRE band. An asinh transfer function is used to maximize the dynamic range of visible structures, and a common field of view of approximately $\sim$$100\times100~\rm kpc^2$ is used to facilitate comparisons across the sample and with images in the GOALS {\em Spitzer} atlas in Mazzarella \etal (2017).  Scale bars indicate 10 kpc and the equivalent angular scale as derived from the angular diameter distance in Table \ref{tbl:BasicGoalsData}.  The beam sizes at each wavelength are indicated on the lower right of each panel.  The atlas is ordered by increasing right ascension.}

\ifnum\Mode=0 
\vskip 0.3in
\placefigure{fig:atlas}
\vskip 0.3in
\else
\ifnum\Mode=2 \onecolumn \fi 

\clearpage
\voffset        -1.0in 

\begin{figure}[!htb]
\figurenum{3}
\label{fig:atlas}
\includegraphics[scale=0.865,angle=0]{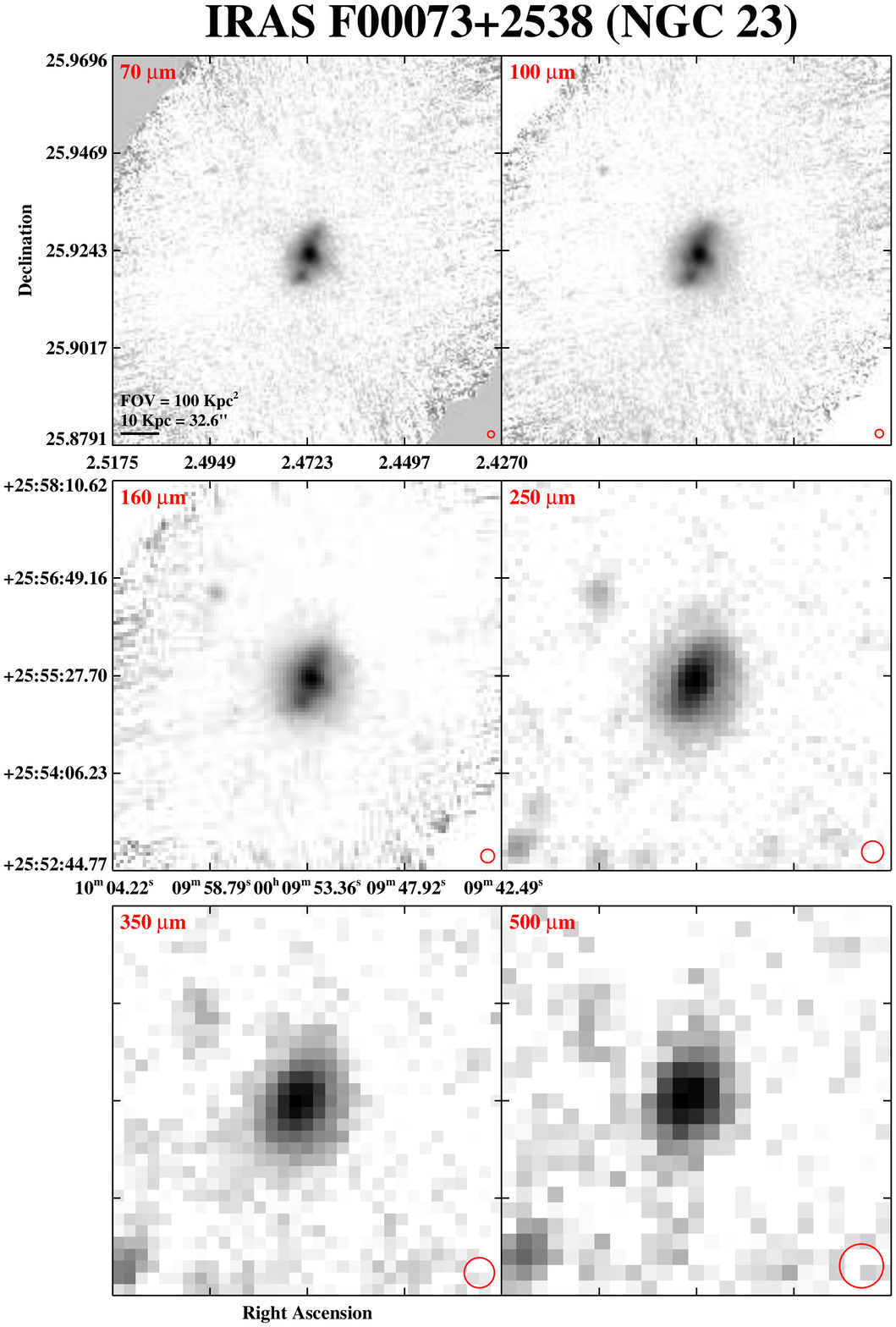}
\caption{\figcapAtlasOne}
\end{figure}
\clearpage

\ifnum\InsertAllPartOne>0  
\newcount\a \a=2    
\newcount\b \b=210   
\newcount\c \c=6    
\newcount\d \d=209 
\newcount\e \e=209

\loop \ifnum\a<\b  
  {
    \voffset       -1.0in
    \begin{figure}
    \figurenum{3}
       \includegraphics[scale=0.865,angle=00]{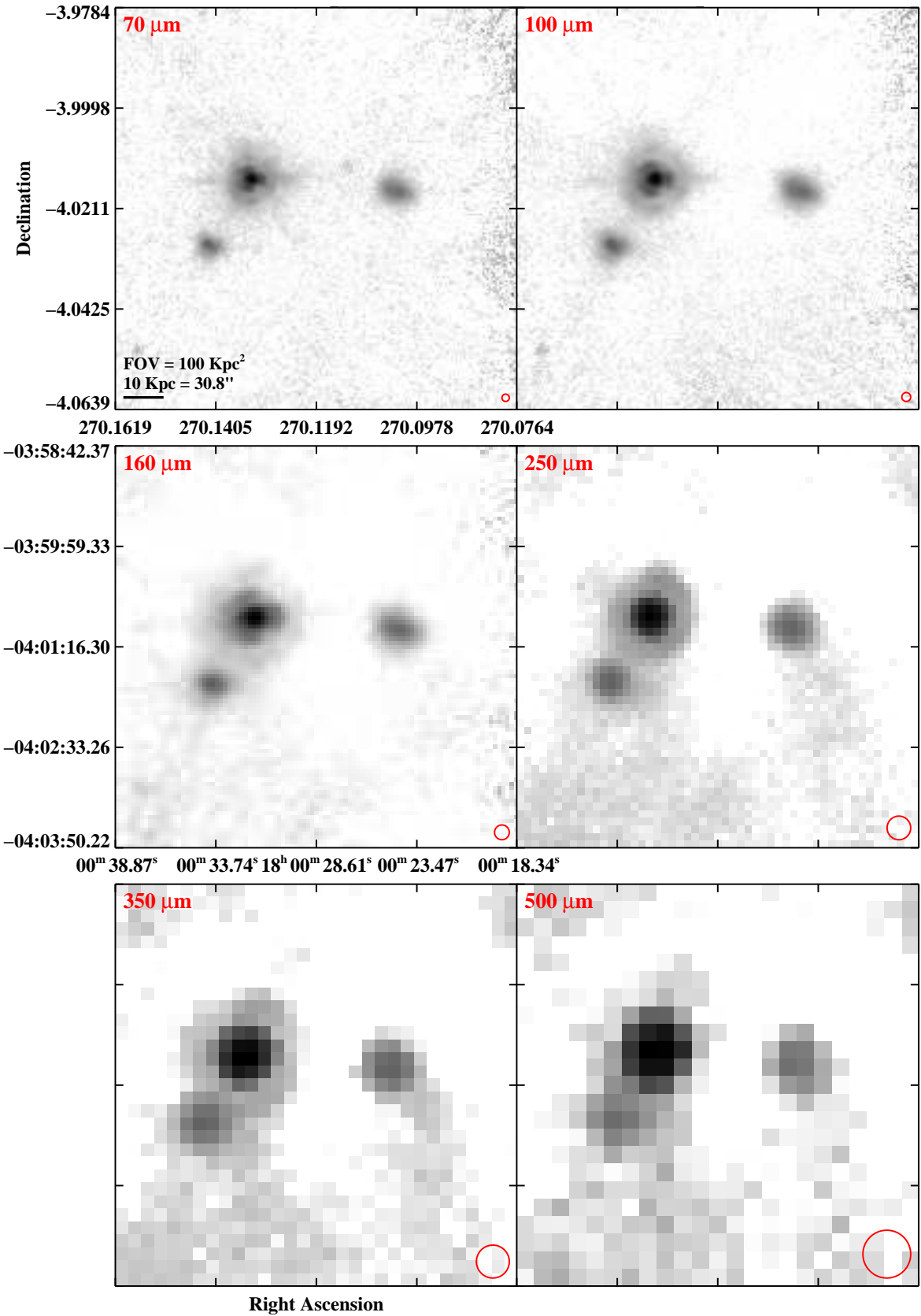}
       \caption{Continued (page \number\a \ of \number\d).}
    \end{figure}
  }
   \clearpage 
   \advance \a 1 
\repeat

\else 
\vskip 5.0in
\fbox{\bf Pages 2 - 209 are omitted here to keep the PDF manageable.

      See {\it http://slirgs.agns.org/}.
      }
\clearpage
\fi 

\ifnum\Mode=2 \twocolumn \fi 
\fi 

\voffset 0.0in 


\ifnum\Mode=0
\placefigure{fig:CurveofGrowthExamples}
\else
\begin{figure*}[tb*]
\figurenum{4}
\label{fig:CurveofGrowthExamples}
\center
\vskip -0.6in
\includegraphics[scale=0.25,angle=0]{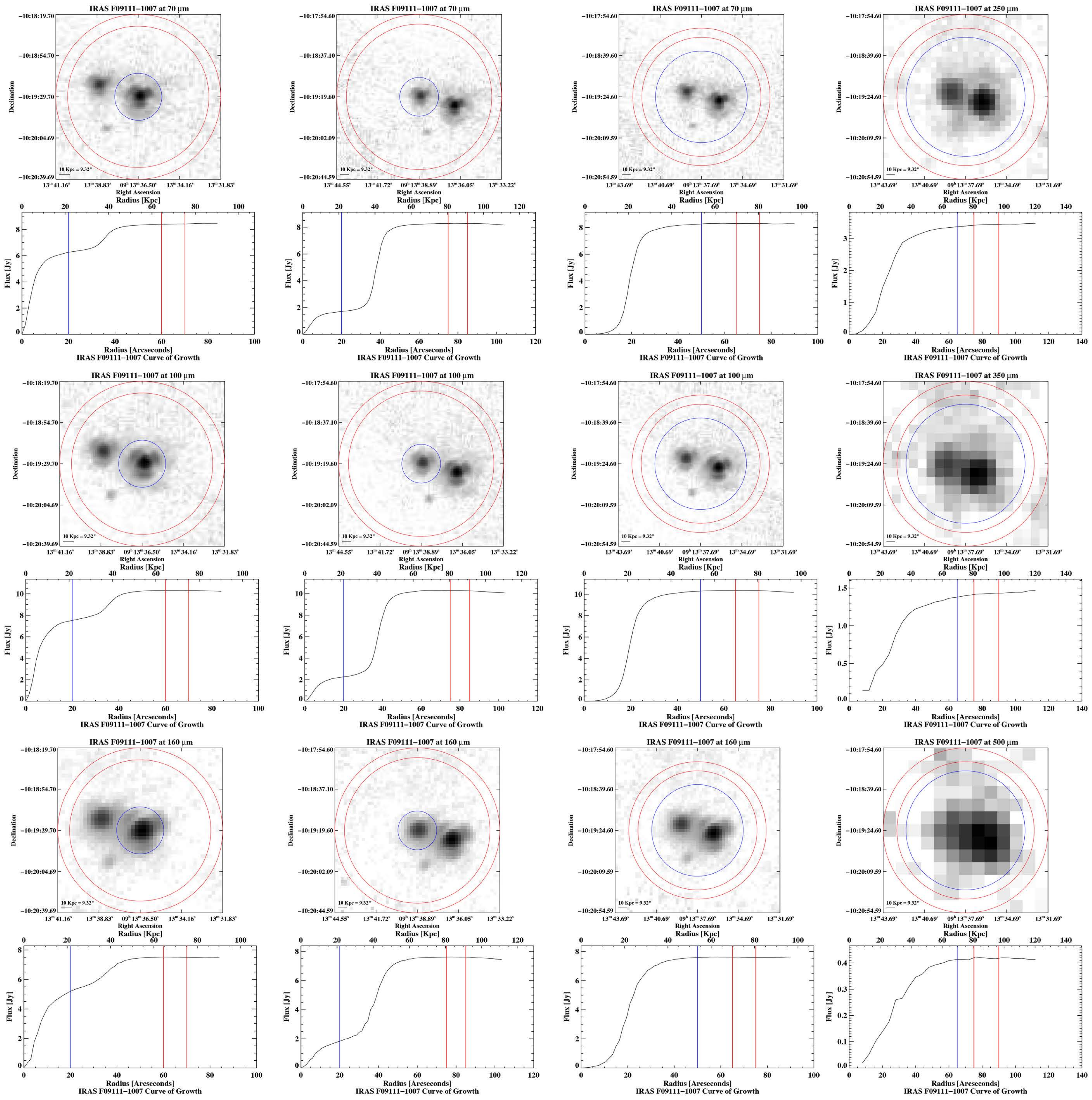}
\caption{Twelve curve of growth plots for IRAS F09111--1007, which are representative for the entire GOALS sample.  The blue circle in each image is the photometry aperture, while the red circles are the annuli from which the background is measured.  These circles are represented in the curve of growth plot immediately below each image.  The first column shows the PACS 70 \mum, 100 \mum, and 160 \mum~photometry apertures for the western component of the system.  The second column shows the PACS photometry apertures for the eastern nucleus.  The third column shows the PACS photometry apertures encompassing both galaxies which includes flux not in the component apertures, giving the total flux from this system.  In the SPIRE bands, we only computed component fluxes at 250 \mum~since the galaxy pair is still resolved, however since the galaxies are essentially unresolved in the other two SPIRE bands, we only compute total fluxes at those two wavelengths.  Note the fourth column only shows the total SPIRE apertures of both galaxies, and the individual 250 \mum~plots were omitted to keep the figure manageable.}
\end{figure*}
\fi

\section{{\em Herschel}-GOALS Aperture Photometry}
In this section we discuss the manner in which the broadband photometry were determined for our sample.  Both PACS and SPIRE photometry were obtained using the {\tt annularSkyAperturePhotometry} routine found in HIPE.  At first we attempted to measure fluxes by using an automated routine to determine the appropriate circular aperture sizes for each galaxy, based on data from the MIPS instrument on {\em Spitzer}.  Unfortunately this approach does not work well for our sample, due to the extended nature of some GOALS systems and galaxies.

Instead we concluded that the best approach was to determine apertures by visual inspection, and subsequently check that we included all of the flux by plotting a curve of growth.  We found that after subtracting any offset in the background levels, the curve of growth almost always flattens out at large radii, indicating a background flux contribution of zero.  There are only a few small cases in the PACS data where the curve of growth does not flatten out, and in all cases this occurs when the object is very faint ($F_\lambda \lesssim0.5$ Jy) and the background noise is more dominant.  Curve of growth plots for the SPIRE data are also flat at large radii even for faint fluxes, again indicating robust background subtractions.  In Figure \ref{fig:CurveofGrowthExamples} we show a set of representative curve-of-growth plots and apertures for IRAS F09111--1007 at different wavelengths.  The photometry aperture is represented by the blue circle in the image and the blue line in the curve of growth plot below it.  In order to facilitate comparison of matched aperture fluxes, all PACS aperture sizes are identical, while all SPIRE apertures are also identical, but larger than that of PACS.  The aperture radius is typically set by the band with the largest beam size in which we can make a measurement for each instrument, which is usually the 160 \mum~channel for PACS and the 500 \mum~channel for SPIRE.  We found that aperture radii encompassing approximately 95\% of the total light gave the best tradeoff between including all of the flux but at the same time keeping the background error from getting too high.  Although it is possible to use the same aperture size across all six bands (\ie the SPIRE aperture size), the larger SPIRE aperture would encompass a significant amount of sky background for the higher resolution images (\ie at 70 \mum) and would introduce additional noise in our measurements.  We therefore decided it was best to match the apertures for each instrument.

To accurately measure the flux of each galaxy the sky background must be subtracted from the measured flux.  To do this we estimate the sky background in the annulus represented by the red circles in the image, which corresponds to the two red lines in the curve of growth plot.  These background annuli were chosen to be as free from any source emission as possible.  Within the {\tt annularSkyAperturePhotometry} routine we used the sky estimation algorithm from {\tt DAOPhot}\footnote{Adapted from the IDL AstroLib {\tt mmm.pro} routine.} to estimate the sky level, with the ``fractional pixel" setting enabled.  The background corrected flux density is then the total flux minus the product of the measured background level and the number of pixels within the target aperture.

We note that in some cases both component and total fluxes are measured for close pairs.  These galaxies can be easily resolved and separated at shorter wavelengths, but become unresolved at longer wavelengths.  In order to choose the best flux aperture, we carefully selected the radius at which the curve of growth was flattest.  This is apparent in Figure \ref{fig:CurveofGrowthExamples} in the first two columns where the galaxy pair is easily resolved at 70 \mum, but becomes marginally resolved at 160 \mum.  The third column in Figure \ref{fig:CurveofGrowthExamples} shows the curve of growth from a single large aperture encompassing the entire system, which includes faint extended flux missed by the individual component apertures.  Finally since the galaxy pair is unresolved in the 350 \mum~and 500 \mum~SPIRE bands, we do not measure any component fluxes at those wavelengths.  At 250 \mum, component fluxes are still computed since they pair is still resolved.  Every effort was made to measure as many marginally resolved systems as possible, while also providing a total flux measurement from one large aperture when necessary.  We believe separately measuring component and total fluxes in cases such as this will be useful when the fractional flux contribution of each component is desired.

In Table \ref{tbl:TotalFluxes} we present the table of monochromatic total flux density for each GOALS system in units of Jansky.  Depending on the number of galaxies within a system, their apparent separation on the sky, and the beam size at that particular wavelength, the total {\em Herschel} flux for each system is calculated using one of three methods.  In the simple case of single galaxy, the system flux is just the flux of that galaxy.  In cases where there are two or more galaxies that are widely separated, the total flux is the sum of the component fluxes measured in separate apertures.  Finally in cases where component galaxies are resolved but still overlapping (\ie in Figure \ref{fig:CurveofGrowthExamples}), the total system flux is obtained from a single large aperture encompassing all of the components.  For triple and quadruple systems where two galaxies are close and a third (or fourth) is far away (\ie IRAS F02071--1023), the total system flux is calculated using a hybrid method: a single large aperture for the two close components, plus a second (or third) aperture around the far component(s).  Since all of the total fluxes are calculated differently, we omit the coordinate and aperture radius in Table \ref{tbl:TotalFluxes}, but we include it in Table \ref{tbl:ComponentFluxes} (see below).

In Table \ref{tbl:TotalFluxes}, Column (1) is the row reference number (corresponding to Tables 1, 2, and 4) while column (2) is the IRAS name of the galaxy, ordered by ascending RA.  Galaxies with the ``F" prefix originate from the {\em IRAS} Faint Source Catalog, and galaxies with no ``F" prefix are from the Point Source Catalog.  Column (3) lists common optical counterpart names to the galaxy systems.  Columns (4) -- (6) are the total fluxes from the PACS instrument in units of Jy.  Note that the four galaxies which lack $100$ \mum~measurements are IRAS F02401-0013 (NGC 1068), IRAS F09320+6134 (UGC 05101), IRAS F15327+2340 (Arp 220), and IRAS F21453-3511 (NGC 7130).  Columns (7) -- (9) are the total fluxes from the SPIRE instrument in units of Jy.

In Table \ref{tbl:ComponentFluxes} we present the table of monochromatic flux density in units of Jansky for each component measurable within each system, with the total system flux from Table \ref{tbl:TotalFluxes} included for completeness on the last line for each system.  For total fluxes that do not have an aperture size listed, the totals were calculated as the sum of the components.  Likewise the RA and declination for these systems (on the totals line) represent the geometric midpoint between the companion galaxies.  The column descriptions are (1) the row reference number, which corresponds to the same indices used in Tables 1--3.  Column (2) is the IRAS name of the galaxy, ordered by ascending RA.  Column (3) is the individual name to that galaxy component.  Note that galaxies prefixed by IRGP are from the catalog of newly defined infrared galaxy pairs defined in the companion {\em Spitzer}-GOALS paper by Mazzarella \etal (2017).  Columns (4) -- (5) are the coordinates of the aperture centers used.  Lines where coordinates are listed but have no aperture radii are cases where the total flux is the sum of two widely separated components.  These are the same 8 \mum~coordinates adopted in Mazzarella \etal (2017), however a few were slightly adjusted for the {\em Herschel} data.  Columns (6) -- (7) are the aperture radii used for PACS photometry, in arcsec and kpc respectively.  Columns (8) -- (10) are the fluxes from the PACS instrument in units of Jy.  Galaxy components that do not have flux measurements are too close to a companion galaxy to be resolved by PACS.  Columns (11) -- (12) are the aperture radii used for SPIRE photometry, in arcsec and kpc respectively.  Finally columns (13) -- (15) are the fluxes from the SPIRE instrument in units of Jy.  Galaxy components that do not have flux measurements are too close to a companion galaxy to be resolved by SPIRE.

\subsection{PACS Aperture Photometry}

In addition to measuring the flux, we must apply an aperture correction to account for flux outside of the aperture.  The PACS aperture corrections are determined from observations of bright celestial standards, and the correction factors are included in the PACS calibration files distributed from the HSA.  Within HIPE, the {\tt photApertureCorrectionPointSource} task performs the aperture correction, where the input is the output product from the aperture photometry task.  In addition a responsivity version must be specified, which for our data we used the most recent version (FM 7\footnote{For a description, see section 2.3 of the PACS calibration framework document:
http://herschel.esac.esa.int/twiki/pub/Pacs/PacsCalibration/
The\_PACS\_Calibration\_Framework\_-\_issue\_0.13.pdf}).  Since these aperture corrections are only applicable to point sources at each wavelength, we only apply the aperture correction to point sources within our sample.  To identify the point sources, we performed PSF fitting of each source in our sample, and selected the objects with FWHM consistent with the corresponding point source FWHM in each PACS band.  In Table \ref{tbl:ComponentFluxes} we denote the fluxes in which an aperture correction was applied by the superscript c.  Typical (average) aperture correction values for the 70, 100, and 160 \mum~bands are 11.7\%, 12.8\%, and 15.5\% of the uncorrected flux, respectively.  The median values of the aperture correction values are less than a percent away from the averages.  We do not flag aperture-corrected fluxes in Table \ref{tbl:TotalFluxes} since many of the total fluxes are a combination of aperture-corrected and uncorrected fluxes.

We also experimented with applying these corrections to marginally resolved systems and systems with a point source and extended flux, however we found that the aperture corrections artificially boosted the flux by approximately 6\% on average.  This is because many of our objects have varying levels of flux contribution from the point source and extended component.  Furthermore, the PACS team performed a careful surface brightness comparison\footnote{For more details see the {\em Herschel} technical note PICC-NHSC-TN-029.} of PACS data with that of {\em IRAS} and {\em Spitzer} MIPS data on the same fields.  By convolving, converting, and re-gridding the higher resolution PACS 70 \mum~to that of {\em IRAS} 60 \mum~and MIPS 70 \mum, and the PACS 100 \mum~maps to that of {\em IRAS} 100 \mum~it was shown that there is no need to apply any pixel-to-pixel gain corrections to the PACS data.  They also conclude that their point-source based calibration scheme is applicable in the case for extended sources.  A similar conclusion is reached for the PACS red array\footnote{See technical note PICC-NHSC-TR-034.}.  \citet{2014ApJ...794..152M} also found in their {\em Herschel} PACS observations of the {\em Swift} BAT sample that aperture corrections on extended sources were negligible (less than 3\%).  Therefore we leave sources appearing extended or semi-extended in our sample unaltered by any aperture correction.

The absolute flux calibration of PACS uses models of five different late type standard stars with fluxes ranging between 0.6--15 Jy in the three photometric bands \citep{2014ExA....37..129B}.  In addition, ten different asteroids are also used to establish the flux calibration over the range of 0.1--300 Jy \citep{2014ExA....37..253M}.  For the standard stars, the absolute flux accuracy is within 3\% at 70 \mum~and 100 \mum, and within 5\% at 160 \mum.  In addition, Uranus and Neptune were also observed for validation purposes with fluxes of up to several hundred Jy, however a 10\% reduction due to nonlinearity in the detector response was observed.  Taken altogether, the error in flux calibration is consistent to within 5\% of the measured flux and takes into account flat-fielding, responsivity correction which includes the conversion of engineering units from volts to Jy pixel$^{-1}$, and gain drift correction which corrects for small drifts in gain with time (PACS Observer's Manual, and references therein).  Since PACS did not perform absolute measurements over the course of the mission, the fluxes are only measured relative to the zero level calculated by the mappers which is arbitrary.

In addition to the flux calibration uncertainty, we must also take into account the error from the background subtraction as well as the instrumental error.  The error from the background subtraction is calculated in the following manner: first using the HIPE implementation of {\tt DAOPhot} the 1-$\sigma$ dispersion is calculated from all the pixels within the background annulus surrounding the target aperture.  This is then multiplied by the square root of the total number of pixels within photometry aperture, under the assumption that the error in background subtraction of individual pixels are not correlated.  On the other hand the instrumental error is calculated as the quadrature sum of all error pixels within the target aperture, using the error maps produced by the mapmaker.  The total flux uncertainty is then calculated as the quadrature sum of all three error components.

We note that only two of the three galaxies in IRAS F07256+3355 (NGC 2388) were observed by PACS due to the smaller field of view, while SPIRE observed all three.  Consequently the total fluxes in Table \ref{tbl:TotalFluxes} for this system is the sum of only the two galaxies observed by both instruments, however SPIRE photometry of the third galaxy to the west is provided in the component flux table (Table \ref{tbl:ComponentFluxes}).  The same is also true for IRAS F23488+1949, with the third galaxy to the NNW of the closer pair.

\def\tableTotalFluxes{
\begin{deluxetable}{cllccccccc}
\tabletypesize{\footnotesize} 
\setlength{\tabcolsep}{0.06in} 
\rotate
\singlespace
\tablenum{3}
\tablewidth{0pt}
\tablecaption{PACS and SPIRE Total Fluxes of GOALS Systems}\label{tbl:TotalFluxes}

\tablehead{
\multicolumn{1}{c}{} &  
\multicolumn{1}{c}{} &
\multicolumn{1}{c}{} &
\multicolumn{3}{c}{PACS}   &
\multicolumn{1}{c}{} &  
\multicolumn{3}{c}{SPIRE} \\
 
\cline{4-6} \cline{8-10} \\
\multicolumn{1}{c}{\#} &  
\multicolumn{1}{c}{IRAS Name} &
\multicolumn{1}{c}{Optical Name} &
\colhead{$\mathrm{F}_{\lambda}(70~\mu\mathrm{m})$} &
\colhead{$\mathrm{F}_{\lambda}(100~\mu\mathrm{m})$} &
\colhead{$\mathrm{F}_{\lambda}(160~\mu\mathrm{m})$} &
\colhead{} & 
\colhead{$\mathrm{F}_{\lambda}(250~\mu\mathrm{m})$} &
\colhead{$\mathrm{F}_{\lambda}(350~\mu\mathrm{m})$} &
\colhead{$\mathrm{F}_{\lambda}(500~\mu\mathrm{m})$} \\

\colhead{} &
\colhead{} &
\colhead{} &
\colhead{Jy} &
\colhead{Jy} &
\colhead{Jy} &
\colhead{} & 
\colhead{Jy} &
\colhead{Jy} &
\colhead{Jy} \\

\colhead{(1)} &     
\colhead{(2)} &     
\colhead{(3)} &     
\colhead{(4)} &     
\colhead{(5)} &     
\colhead{(6)} &     
\colhead{} &          
\colhead{(7)} &    
\colhead{(8)} &    
\colhead{(9)}       
}
\startdata
1 & F00073+2538 & NGC 23 & $10.48\pm0.52$ & $14.26\pm0.71$ & $11.85\pm0.59$ & ~ & $5.311\pm0.347$ & $2.026\pm0.134$ & $0.718\pm0.046$\\
2 & F00085-1223 & NGC 34, Mrk 938 & $18.21\pm0.84$ & $17.68\pm0.81$ & $10.65\pm0.48$ & ~ & $3.573\pm0.214$ & $1.239\pm0.075$ & $0.339\pm0.021$\\
3 & F00163-1039 & Arp 256, MCG-02-01-051/2 & $8.038\pm0.403$ & $9.876\pm0.495$ & $7.496\pm0.375$ & ~ & $2.862\pm0.189$ & $1.209\pm0.081$ & $0.419\pm0.033$\\
4 & F00344-3349 & ESO 350-IG 038, Haro 11 & $6.195\pm0.276$ & $5.096\pm0.228$ & $2.341\pm0.101$ & ~ & $0.639\pm0.038$ & $0.216\pm0.014$ & $0.067\pm0.006$\\
5 & F00402-2349 & NGC 232 & $14.60\pm0.56$ & $20.00\pm0.77$ & $16.11\pm0.60$ & ~ & $6.397\pm0.323$ & $2.441\pm0.124$ & $0.703\pm0.035$\\
6 & F00506+7248 & MCG+12-02-001 & $25.31\pm1.27$ & $29.07\pm1.45$ & $20.34\pm1.02$ & ~ & $7.968\pm0.485$ & $2.793\pm0.174$ & $0.896\pm0.057$\\
7 & F00548+4331 & NGC 317B & $10.91\pm0.55$ & $13.07\pm0.65$ & $9.945\pm0.498$ & ~ & $4.014\pm0.266$ & $1.600\pm0.108$ & $0.478\pm0.039$\\
8 & F01053-1746 & IC 1623, Arp 236 & $25.56\pm1.28$ & $29.28\pm1.47$ & $21.46\pm1.08$ & ~ & $8.991\pm0.546$ & $3.277\pm0.201$ & $1.023\pm0.062$\\
\enddata
\tablecomments{
The monochromatic flux density in units of Jansky for each of the three PACS and three SPIRE broadband filters.  These are the total fluxes for each GOALS system.
The column descriptions are
(1) the row reference number.
(2) The IRAS name of the galaxy, ordered by ascending RA.  Galaxies with the ``F" prefix originate from the {\em IRAS} Faint Source Catalog, and galaxies with no ``F" prefix are from the Point Source Catalog.
(3) Common optical counterpart names to the galaxy systems.
Columns (4) -- (6) are the fluxes from the PACS instrument in units of Jy.  Note that the four galaxies which lack $100$ \mum~measurements are IRAS F02401-0013 (NGC 1068), IRAS F09320+6134 (UGC 05101), IRAS F15327+2340 (Arp 220), and IRAS F21453-3511 (NGC 7130).
Columns (7) -- (9) are the fluxes from the SPIRE instrument in units of Jy.
}
\tablenotetext{a}{These are very widely separated galaxy pairs that required two {\em Herschel} PACS observations.}
\tablenotetext{b}{This is a triple system, however only two component galaxies are visible in PACS due to its smaller field of view.  The total flux for this system does {\em not} include the third galaxy not visible in PACS.}
\end{deluxetable}
}

\clearpage
\ifnum\Mode=0 
\vskip 0.3in
\vskip 0.3in
\placetable{tbl:no3}
\else
\ifnum\Mode=2 \onecolumn \fi 
\tableTotalFluxes
\ifnum\Mode=2 \twocolumn \fi 
\fi 

\subsection{SPIRE Aperture Photometry}
The SPIRE 2-pass pipeline (see \S4.2.2) produces a point-source calibrated map as the main output.  However since many of our objects appear extended or marginally extended in our sample, and following the recommendation from the NASA Herschel Science Center (NHSC), we opted to use the extended-source calibrated maps from which we measured all of the fluxes.  Both sets of maps are produced nearly identically, however the extended-source calibrated maps have relative gain factors applied to each bolometer's signal, which accounts for the small differences between the peak and integral of each individual bolometer's beam profile.  This method helps reduce residual striping in maps with extended sources, since the relative photometric gains between all of the bolometers is properly accounted for.  In addition to applying the relative gains, the PMW and PLW channels are zero-point corrected by applying a constant offset based on the {\em Planck}-HFI maps (see \S 6.2.1).  The overall calibration scheme for point and extended sources is described in \citet{2013MNRAS.434..992G}.

The primary flux calibrator for SPIRE is Neptune, chosen because it has a well-understood submillimeter/FIR spectrum and is essentially a point source in the SPIRE beams.  It is also bright enough from which high signal-to-noise measurements can be made, but not so bright that it would introduce non-linearity effects from the instrument.  In order to calibrate the entire instrument, special `fine scan' observations were taken such that each bolometer was scanned across Neptune in order to absolutely calibrate each bolometer.  Repeated observations of Neptune also showed that there were no statistically significant changes in the detector responses over the mission.  Further details on using Neptune as the primary SPIRE flux calibrator can be found in \citet{2013MNRAS.433.3062B}.

Since the vast majority of our sources have fluxes above 30 mJy, \citet{2014ExA....37..175P} recommends using either the timeline fitter or aperture photometry.  Because a significant fraction of our sample contains marginally to very extended sources, as well as point sources, we opted to measure all of our SPIRE fluxes using the {\tt annularSkyAperturePhotometry} task in HIPE in order to keep our measurements as uniform as possible.  However this method results in the loss of flux outside the finite-sized aperture, for which an aperture correction is needed to fully account for all the flux.  In the case of point sources, we applied the aperture correction by dividing our fluxes by the encircled energy fraction (EEF) amount corresponding to the aperture radius and SPIRE channel.  The EEFs can be found in the SPIRE calibration files (accessible from within HIPE), and represents the ratio of flux (energy) inside the aperture divided by the true flux of the point source.  As with the PACS aperture corrections, SPIRE fluxes in which an aperture correction was applied are denoted by a superscript c in Table \ref{tbl:ComponentFluxes}, with average corrections of 10.1\%, 10.3\%, and 14.8\% for the 250, 350, and 500 \mum~channels respectively.  Similarly the median correction values are less than a percent difference from the averages.

In order to check the validity of our point source fluxes, we measured our fluxes a second time using the timeline source fitter on a subset of 65 objects that are point sources in all three SPIRE bands.  The timeline fitter is the preferred method of obtaining point source fluxes on the SPIRE maps, since it works on the baseline subtracted, destriped, and deglitched Level 1 timelines of the data (which are calibrated in Jy/beam\footnote{See \cite{2010SPIE.7731E..36D}, \S5.}).  By using a Levenberg-Marquardt algorithm to fit a two dimensional circular or elliptical Gaussian function to the 2-D timeline data, the source can be modeled and the point source flux can be calculated from the 2-D fit.  The advantage is it avoids any potential artifacts arising from the map-making process, such as smearing effects from pixelization.  Because it does not use the Level 2 maps, source extraction is not necessary (\ie aperture photometry), and there are no aperture corrections needed since the 2-D fit in principle takes into accounts all of the flux from the point source.

When we compared the aperture photometry results to the timeline fitter results, we found that they both agree very well at 250 \mum~and 350 \mum~with an average aperture/timeline flux ratio of 1.030 and 0.995 respectively, however the 500 \mum~channel had a slightly lower ratio of 0.93.  To further check our results, we plotted the aperture/timeline flux ratio against the aperture photometry flux for all three bands, and found no statistically significant correlation in the flux ratio as a function of flux.  However we do note in the 500 \mum~case, fluxes less than approximately 150 mJy appear to have a lower aperture/timeline flux ratio, whereas fluxes above that value have an average ratio close to unity.  We believe this underestimation at faint fluxes is due to confusion noise, which was also observed in the SPIRE Map Making Test Report.  Furthermore we note that the discrepancy in the 500 \mum~fluxes are still consistent within the typical flux errors ($\sim$$15$\%).  As a final check we also plotted the aperture/timeline ratio against the aperture photometry radius, and we again found no statistically significant correlation.  From these tests our point source aperture photometry fluxes appear to be in good agreement with the results from the timeline fitter.

In the case of semi-extended to extended sources, aperture corrections become more complex since the flux originates not from an unresolved source, but is seen instead as surface brightness distributed within an aperture.  Although an aperture correction is needed for reasons similar to the point source case, \citet{2016MNRAS.456.3335S} found that their extended SPIRE fluxes for their {\em Swift} BAT sample did not need aperture corrections because they were negligible.  To test this, they first convolved their 160 \mum~PACS data to the resolution of the three SPIRE bands, and then measured the fluxes on both the convolved and unconvolved images using the same SPIRE aperture sizes.  Aperture corrections were then calculated as the ratio of the flux on the original PACS image divided by the flux obtained on the convolved image, with resulting median aperture corrections of 1.01, 0.98, and 0.98 for the 250 \mum, 350 \mum, and 500 \mum~channels respectively.  This makes the assumption that the 160 \mum~and SPIRE fluxes originate from the same material within their galaxies.  We also note that their aperture sizes are similar to ours, since their galaxy sample lies in the same redshift range.  \citet{2012A&A...543A.161C} also showed by simulating in the worst-case scenario, a maximum aperture correction of 5\% is needed at 500 \mum.  However this was done on an (intentionally) unphysical source that has a flat constant surface brightness, with a sharp drop to zero flux at a set radius.  On more realistic sources they calculated aperture corrections of approximately $\lesssim$2\%.  As these corrections are very close to unity, we follow their precedent in only reporting the integrated, background subtracted flux for our extended sources.

To calculate the flux uncertainty for the SPIRE photometry, we follow a similar prescription we used for PACS.  The first is a systematic error in the flux calibration related to the uncertainty in the models used for Neptune, which is the primary calibrator for SPIRE.  These uncertainties, which are correlated across all three SPIRE bands, are currently quoted as 4\%. The other source of uncertainty is a random uncertainty related to the ability to repeat flux density measurements of Neptune, which is 1.5\% for all three bands.  Altogether, these two sources of uncertainty are added linearly for a total of 5.5\% error in the point source flux calibration.  However in the case of extended emission calibration, there is an additional error of 1\% due to the current uncertainty in the measured beam area that is also added linearly.  This error was recently improved from 4\% with the release of the SPIRE calibration version 14\_2.  Therefore the total uncertainty in the extended source calibration scheme amounts to 6.5\% of the background subtracted flux \citep{2013MNRAS.433.3062B}.

To calculate our total flux uncertainties, we must also include any errors incurred from measuring and subtracting the background from the measured flux, as well as the instrumental error.  To estimate the uncertainty from the background subtraction, we measure the 1-$\sigma$ dispersion of the flux in each pixel within the annular area used for our background measurements.  This is then multiplied by the square root of the number of pixels within the photometry aperture (which can be a fractional amount) to obtain the error in background measurement.  The instrumental error is calculated by summing in quadrature the pixels within the aperture on the error map generated by the pipeline.  We note this underestimates the error because the noise is correlated between pixels.  Our final SPIRE flux uncertainties are then computed as the quadrature sum of all three sources of error.  In the case where the total system flux is the sum of two (or more) components, the flux uncertainty is the quadrature sum of each galaxy component's flux error.

\subsubsection{SPIRE Zero-Point Correction}
Due to the large radiative contribution of {\em Herschel}'s optical components (230 Jy, 250 Jy, 270 Jy for the PSW, PMW, PLW channels respectively), SPIRE can only measure the relative flux on the sky, \ie the flux of the target minus the background level.  During data reduction the SPIRE maps are generated such that the background is approximately normalized to zero, which makes it impossible to determine the absolute flux of the target.  However to recover the absolute flux we used the all-sky maps from the {\em Planck} mission (modified to have a spatial resolution of 8\arcmin~FWHM), since the Planck-HFI 857 GHz and 545 GHz filters match fairly well to the {\em Herschel} 350 \mum~ and 500 \mum~ band passes respectively (see Fig.~5.16 in the SPIRE handbook).  These corrections become more important in sources with very extended flux, since some of the diffuse low surface-brightness flux may be subtracted out.

\citet{2016A&A...588A.107B} performed an in-depth analysis of SPIRE and HFI data on the same fields, and found a very high degree of linearity between the two datasets, as well as a good agreement in the relative calibrations between the two instruments.  The zero-points of the {\em Planck} maps are derived assuming that the zero-point of the Galactic emission can be defined as zero dust emission for a null HI column density\footnote{See the Explanatory Supplement to the Planck 2013 results: {\tt \url{ http://wiki.cosmos.esa.int/planckpla/index.php/CMB\_
and\_astrophysical\_component\_maps\#Thermal\_dust\_emission}}}.  The final step is to apply a slight gain correction to the {\em Planck} maps, which for our data we used the NHSC recommended gain factors of  0.989 and 1.02 for the 857 GHz and 545 GHz channels, respectively.  The {\em Planck} calibration uncertainty for both channels is 10\%.  Using the all-sky {\em Planck} data, zero-point corrections are applied as flux offsets over the entire SPIRE map, and do not affect the SPIRE flux calibrations (which is background subtracted).  We note that these zero point corrections were only applied to the 350 \mum~and 500 \mum~channels only, and the 250 \mum~maps were not corrected since there is no overlap with {\em Planck}.

\def\tableComponentFluxes{
\begin{deluxetable}{cllcccccccccccccc}
\tabletypesize{\scriptsize} 
\setlength{\tabcolsep}{0.03in} 
\rotate
\singlespace
\tablenum{4}
\tablewidth{0pt}
\tablecaption{PACS and SPIRE Total and Component Fluxes of GOALS Systems}\label{tbl:ComponentFluxes}

\tablehead{
\multicolumn{1}{c}{} &  
\multicolumn{1}{c}{} &
\multicolumn{1}{c}{} &
\multicolumn{2}{c}{Aperture Center Coordinate} &
\multicolumn{1}{c}{} &  
\multicolumn{5}{c}{PACS}   &
\multicolumn{1}{c}{} &  
\multicolumn{5}{c}{SPIRE} \\
 
\cline{4-5} \cline{7-11} \cline{13-17} \\
\multicolumn{1}{c}{\#} &  
\multicolumn{1}{c}{IRAS Name} &
\multicolumn{1}{c}{Individual Name} &
\colhead{RA} &
\colhead{Dec.} &
\colhead{} & 
\colhead{Ang. Ap.} &
\colhead{Phys. Ap.} &
\colhead{$\mathrm{F}_{\lambda}(70~\mu\mathrm{m})$} &
\colhead{$\mathrm{F}_{\lambda}(100~\mu\mathrm{m})$} &
\colhead{$\mathrm{F}_{\lambda}(160~\mu\mathrm{m})$} &
\colhead{} & 
\colhead{Ang. Ap.} &
\colhead{Phys. Ap.} &
\colhead{$\mathrm{F}_{\lambda}(250~\mu\mathrm{m})$} &
\colhead{$\mathrm{F}_{\lambda}(350~\mu\mathrm{m})$} &
\colhead{$\mathrm{F}_{\lambda}(500~\mu\mathrm{m})$} \\

\colhead{---} &
\colhead{---} &
\colhead{---} &
\colhead{HH:MM:SS} &
\colhead{DD:MM:SS} &
\colhead{} & 
\colhead{\arcsec} &
\colhead{kpc} &
\colhead{Jy} &
\colhead{Jy} &
\colhead{Jy} &
\colhead{} & 
\colhead{\arcsec} &
\colhead{kpc} &
\colhead{Jy} &
\colhead{Jy} &
\colhead{Jy} \\

\colhead{(1)} &     
\colhead{(2)} &     
\colhead{(3)} &     
\colhead{(4)} &     
\colhead{(5)} &     
\colhead{} & 
\colhead{(6)} &     
\colhead{(7)} &     
\colhead{(8)} &     
\colhead{(9)} &     
\colhead{(10)} &   
\colhead{} &          
\colhead{(11)} &    
\colhead{(12)} &    
\colhead{(13)} &    
\colhead{(14)} &    
\colhead{(15)}       
}
\startdata
1~ & F00073+2538 & ~NGC 23 & $~00:09:53.36~$ & $~+25:55:27.7~$ & ~ & $45$ & $13.81$ & $10.48\pm0.52$ & $14.26\pm0.71$ & $11.85\pm0.59$ & ~ & $70$ & $21.48$ & $5.311\pm0.347$ & $2.026\pm0.134$ & $0.718\pm0.046$\tablenotemark{c}\\[0.6em]
2~ & F00085-1223 & ~NGC 34 & $~00:11:06.56~$ & $~-12:06:28.2~$ & ~ & $45$ & $17.78$ & $18.21\pm0.84$\tablenotemark{c} & $17.68\pm0.81$\tablenotemark{c} & $10.65\pm0.48$\tablenotemark{c} & ~ & $55$ & $21.73$ & $3.573\pm0.214$\tablenotemark{c} & $1.239\pm0.075$\tablenotemark{c} & $0.339\pm0.021$\tablenotemark{c}\\
2~ & F00085-1223 & ~NGC 35 & $~00:11:10.51~$ & $~-12:01:14.9~$ & ~ & \nodata & \nodata & \nodata & \nodata & \nodata & ~ & \nodata & \nodata & \nodata & \nodata & \nodata\\
2~ & F00085-1223 & ~IRGP J001108.5-120351 & $~00:11:08.54~$ & $~-12:03:51.6~$ & ~ & \nodata & \nodata & $18.21\pm0.84$\tablenotemark{c} & $17.68\pm0.81$\tablenotemark{c} & $10.65\pm0.48$\tablenotemark{c} & ~ & \nodata & \nodata & $3.573\pm0.214$\tablenotemark{c} & $1.239\pm0.075$\tablenotemark{c} & $0.339\pm0.021$\tablenotemark{c}\\[0.6em]
3~ & F00163-1039 & ~MCG-02-01-051 & $~00:18:50.90~$ & $~-10:22:36.7~$ & ~ & $24$ & $12.96$ & $7.501\pm0.329$\tablenotemark{c} & $8.382\pm0.364$\tablenotemark{c} & $5.870\pm0.244$\tablenotemark{c} & ~ & \nodata & \nodata & \nodata & \nodata & \nodata\\
3~ & F00163-1039 & ~MCG-02-01-052 & $~00:18:49.85~$ & $~-10:21:34.0~$ & ~ & $38$ & $20.52$ & $1.206\pm0.062$ & $2.081\pm0.105$ & $1.960\pm0.099$ & ~ & \nodata & \nodata & \nodata & \nodata & \nodata\\
3~ & F00163-1039 & ~Arp 256 & $~00:18:50.37~$ & $~-10:22:05.3~$ & ~ & $70$ & $37.81$ & $8.038\pm0.403$ & $9.876\pm0.495$ & $7.496\pm0.375$ & ~ & $80$ & $43.21$ & $2.862\pm0.189$ & $1.209\pm0.081$ & $0.419\pm0.033$\\
\enddata
\end{deluxetable}
}
\voffset = 0.4in

\clearpage
\ifnum\Mode=0 
\placetable{tbl:no4}
\else
\ifnum\Mode=2 \onecolumn \fi 

\tableComponentFluxes
\ifnum\Mode=2 \twocolumn \fi 
\fi 
\voffset = 0in

\ifnum\Mode=0
\placefigure{fig:FluxHistogram}
\else
\begin{figure*}[h*]
\figurenum{5}
\label{fig:FluxHistogram}
\center
\vskip -1.0in
\includegraphics[scale=0.8,angle=0]{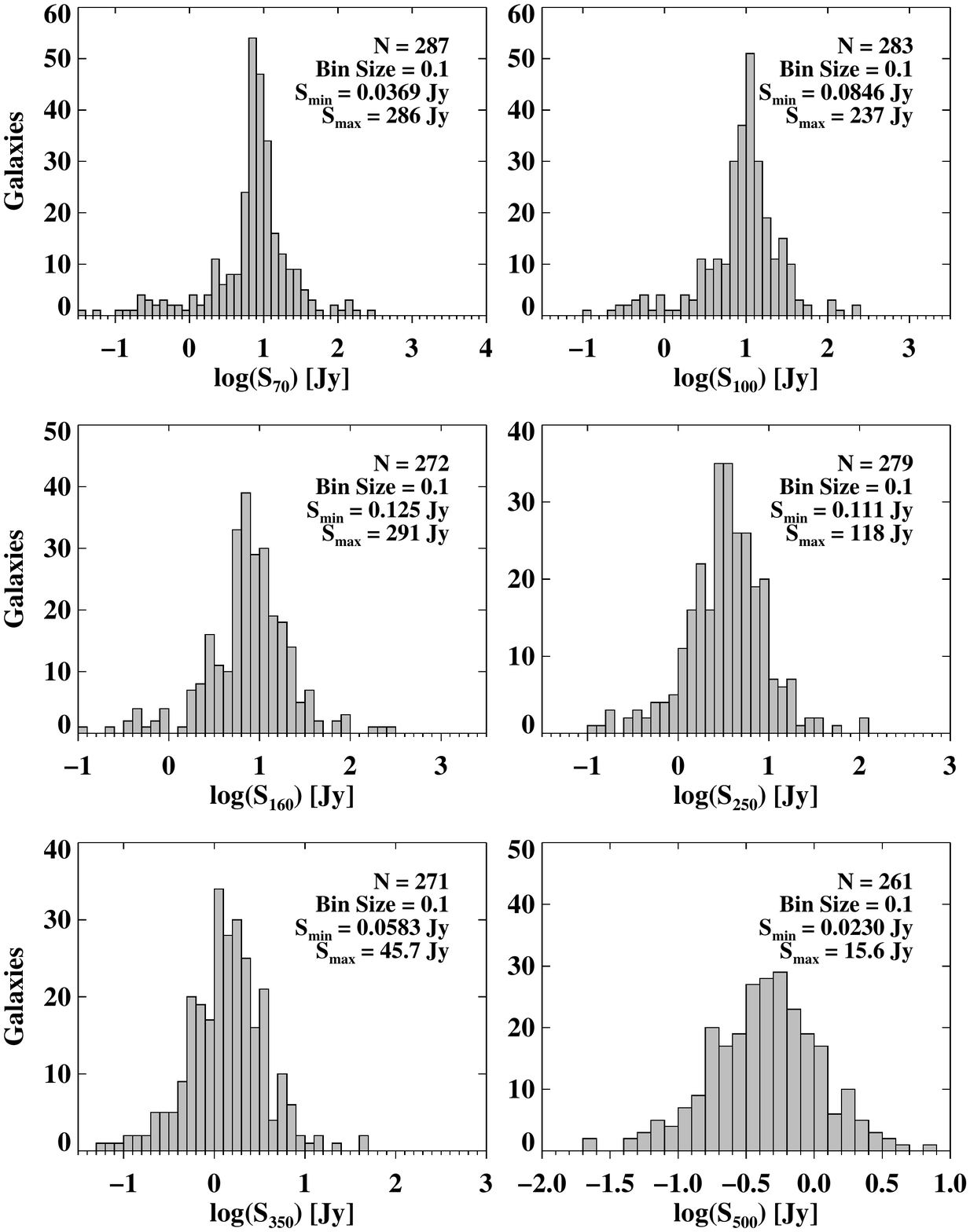}
\caption{Histogram plot of the {\em Herschel} PACS and SPIRE fluxes from our sample.  The histogram range for each band was fine tuned in order to meaningfully show the data.  The fluxes shown here are all the actual measured fluxes, consisting of component and total fluxes.  The $x$-axis of each panel is shown in units of log(Jy) to encompass the wide dynamic range of fluxes measured within the data.}
\end{figure*}
\fi

\subsection{Distribution of {\em Herschel} Fluxes}
In Figure \ref{fig:FluxHistogram} we show the distribution of fluxes from our {\em Herschel} program in each of the three PACS and SPIRE photometer bands.  The histogram $x$-axis range and binning for each band was selected in order to meaningfully show the data.  The fluxes shown here are all 1657 measured fluxes, comprising both component and total fluxes, and do not include total system fluxes that are the sum of the component fluxes.  The $x$-axis of each panel is shown in units of log(Jy) to encompass the wide dynamic range of fluxes measured within the data.

As expected the fluxes are generally higher in the three PACS bands, while they are lower in the SPIRE bands due to the Rayleigh-Jeans tail of the galaxy's SED.  The number of measured fluxes and bin sizes are indicated for each band, as well as the minimum and maximum fluxes.  The galaxies with the highest fluxes are all nearby (IRAS F02401--0013/NGC 1068, IRAS F03316--3618/NGC 1365, and IRAS F06107+7822/NGC 2146) and tend to be quite extended in the {\em Herschel} maps, with the exception of NGC 2146 which appears to be more concentrated than the other two in the PACS 70 \mum~ and 100 \mum~channels.  On the other hand the faintest measured fluxes in the PACS bands are well within the ``faint" flux regime for PACS data reduction (see \S4.1).

\section{Discussion}

\subsection{Comparison of PACS Fluxes to Previous Missions}
One important check is to compare our new PACS 100 \mum~fluxes to the legacy {\em IRAS} 100 \mum~fluxes published in \citet{2003AJ....126.1607S}, since the central wavelengths of both instruments are the same.  In Figure \ref{fig:PacsIrasFilterCurve} we show the filter transmission curves for PACS and {\em IRAS} in blue and red respectively.  Before comparing the fluxes measured from each telescope, several constraints must be used to ensure a meaningful comparison.  Importantly, we only selected objects that either appear as single galaxies in the PACS 100 \mum~maps, or have component galaxies close enough such that it is only marginally resolved (or not at all) by PACS.   We note that the {\em IRAS} 100 \mum~channel has a FWHM beamsize of $\sim$$4$\arcmin, which is significantly larger than the PACS 100 \mum~beamsize of 6\arcs8, therefore any unresolved system in PACS would certainly appear unresolved to {\em IRAS}.  Second, we also applied an aperture correction for point source objects in the PACS 100 \mum~maps, however we did not apply a color correction to any of our fluxes (see \S7.3.1).  The latter point would be needed to stay in accordance with how \citet{2003AJ....126.1607S} measured the {\em IRAS} RBGS fluxes \citep[see also][]{1989AJ.....98..766S}, to ensure as accurate of a comparison as possible\footnote{The {\em IRAS} data reduction pipeline also assumes a power law spectral index of $-1$, which is the same as PACS and SPIRE.}.  Importantly, these objects span the entire range of 100 \mum~fluxes within the GOALS sample, and represent the entire spectrum of source morphology from point source to very extended objects.

In the upper panel of Figure \ref{fig:PacsIrasComparison} we plot the 100 \mum~PACS/IRAS flux ratio as a function of the {\em IRAS} 100 \mum~flux for 128 GOALS objects satisfying our criteria (corresponding to 64\% of our sample).  The red line represents the unweighted average of the ratio which is 1.012, with dashed lines representing the $1$-$\sigma$ scatter of 0.09.  On the other hand the median of the PACS/IRAS ratio is 1.006.  Additionally we see no variation in the flux ratio except for fluxes above $\sim$100 Jy, where our PACS fluxes are slightly higher.  The IRAS names of these six galaxies are F03316--3618, F06107+7822, F10257--4339, F11257+5850, 13242--5713, and F23133--4251.  Of these six sources the two with the highest PACS/IRAS ratios, F03316--3618 (NGC 1365) and F06107+7822 (NGC 2146), are large galaxies with optical sizes of $11\arcm2 \times 6\arcm2$ and $6\arcm0 \times 3\arcm4$.  Their fluxes could be underestimated by IRAS since they were computed assuming point source photometry, however once we exclude these two systems, there doesn't appear to be any PACS excess left in the bright sources.  Overall, there is a broad agreement in fluxes between our {\em Herschel} data and the {\em IRAS} data, to within measurement errors ($\sim$5--10\% for PACS).
 
Additionally we also compared the PACS 70 \mum~fluxes to the {\em IRAS} 60 \mum~fluxes, however because of the difference in wavelength, we first had to interpolate the {\em IRAS} 60 \mum~measurement to 70 \mum.  To do this, we first estimated the power law index to the nearest whole number on the short-wavelength side of the SED bump using the {\em IRAS} 60 \mum~and PACS 70 \mum~fluxes\footnote{We did not use the {\em IRAS} 100 \mum~flux as that is right on the peak of the SED, which would systematically underestimate the power law index.}.  To interpolate the {\em IRAS} 60 \mum~flux to 70 \mum, we divided the {\em IRAS} 60 \mum~fluxes by multiplicative factors corresponding to each power law index found in Table 2 of the {\em Herschel} technical note PICC-ME-TN-038.  These factors were calculated by the PACS team to convert PACS fluxes to other key wavelengths and vice versa based on SED shape.  We then plotted the ratio of the PACS 70 \mum~flux to the interpolated {\em IRAS} 70 \mum~flux as a function of the {\em IRAS} flux, shown in the bottom panel of Figure \ref{fig:PacsIrasComparison}.  The average flux ratio represented by the red line is 1.001 with a $1$-$\sigma$ scatter of 0.04 (dashed lines), and the median ratio is 1.00.  The agreement between the PACS and {\em IRAS} data in this case is exquisite, with an even tighter relation than the 100 \mum~comparison throughout the entire flux range.

Another comparison is to perform a similar analysis using GOALS data from the {\em Spitzer} MIPS instrument at 70 and 160 \mum~(Mazzarella \etal 2017, in prep).  Unfortunately, many of the images from that program suffer from saturation and other image quality issues that make it impossible to draw a meaningful comparison.  As a result we have agreed that the PACS 70 and 160 \mum~data will completely supersede the corresponding MIPS data.

The results here are also similar to the analysis done in the {\em Herschel} technical note SAp-PACS-MS-0718-11, where extended source fluxes were compared between PACS to {\em Spitzer}-MIPS and {\em IRAS}.  Although they found an average PACS/IRAS 100 \mum~flux ratio of 1.32, their dispersion in the flux ratio is very similar to our results in Figure \ref{fig:PacsIrasComparison}.  We note that their analysis was done on HIPE 6, where the PACS responsivity was not well understood resulting in much higher flux ratios than our result.

\ifnum\Mode=0 
\placefigure{fig:PacsIrasFilterCurve}
\vskip 0.3in
\else
\begin{figure}[!htb]
\figurenum{6}
\label{fig:PacsIrasFilterCurve}
\includegraphics[scale=0.3,angle=0]{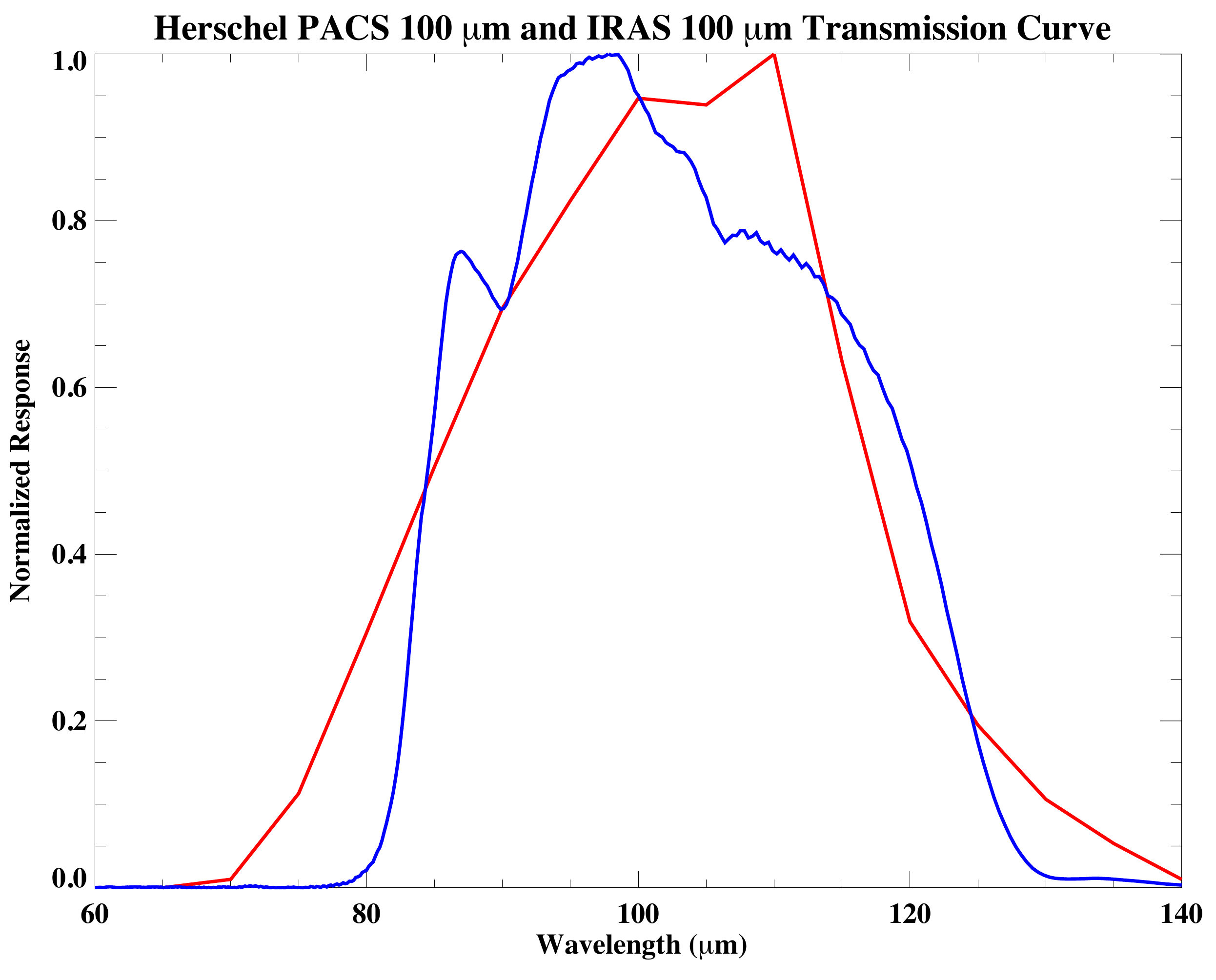}
\caption{The normalized transmission curves of the 100 \mum~band passes for {\em Herschel}-PACS in blue and {\em IRAS} in red.}
\end{figure}
\fi

\ifnum\Mode=0
\placefigure{fig:PacsIrasComparison}
\else
\begin{figure*}[tb*]
\figurenum{7}
\label{fig:PacsIrasComparison}
\center
\vskip -0.6in
\includegraphics[scale=0.7,angle=0]{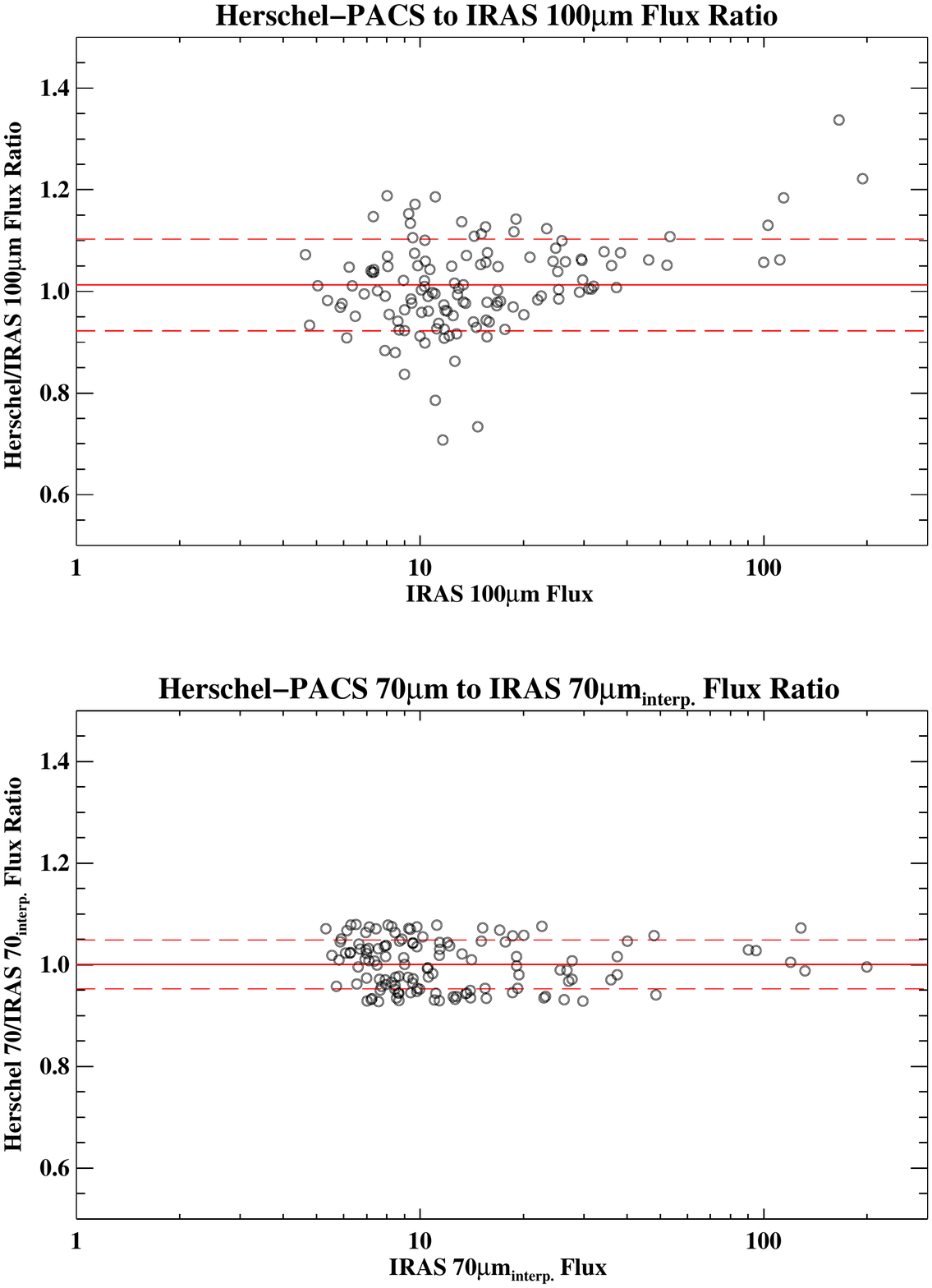}
\caption{{\em Upper panel}: The {\em Herschel}-PACS 100 \mum~to {\em IRAS} 100 \mum~flux ratio plotted as a function of the {\em IRAS} 100 \mum~flux for 128 of our galaxies carefully chosen to be single objects, or if the system has multiple components they are too close to be distinguishable by PACS at 100 \mum.  These galaxies represent the entire spectrum of very extended emission, to point sources as seen by PACS.  The mean ratio represented by the red line is 1.012, with the dashed red lines representing the $1$-$\sigma$ scatter of 0.09.  The median ratio is 1.006.  There appears to be no significant systematic offset, nor is there any evidence of a slope signifying a change in the flux ratio at different {\em IRAS} 100 \mum~flux.  Error bars were omitted to keep the plot readable.  {\em Lower panel}: Same as the upper panel but for the {\em Herschel}-PACS 70 \mum~data compared to the interpolated IRAS 70 \mum~flux.  The mean ratio is 1.001 with a $1$-$\sigma$ scatter of 0.04, and a median ratio of 1.00.  The agreement between the PACS 70 \mum~and interpolated {\em IRAS} 70 \mum~fluxes is excellent.}
\end{figure*}
\fi

\subsection{Comparison of SPIRE Fluxes Measured From Different Calibration Versions}
To check the consistency of our SPIRE fluxes we compared the measured fluxes of our SPIRE data reduced using three different SPIRE calibrations: {\tt SPIRECAL\_10\_1}, {\tt SPIRECAL\_13\_1}, and the latest version {\tt SPIRECAL\_14\_2}.  In Figure \ref{fig:SpireCalComparison} we show six histograms of the fractional percentage change in flux between each calibration version for each of the bands.  In order to facilitate as direct of a comparison as possible, we use the uncorrected fluxes computed directly by the {\tt annularSkyAperturePhotometry} task in HIPE, which are not aperture or color corrected.  The histograms show as a general trend towards longer wavelengths, a larger variance in the percent change in flux.  This is again due to the long wavelength Rayleigh-Jeans tail of the galaxy's SED, where the fainter fluxes are affected more by instrument uncertainties.

In the histogram comparing {\tt SPIRECAL\_10\_1} and {\tt SPIRECAL\_13\_1} (Fig. \ref{fig:SpireCalComparison}, first column), the general trend is an increase in the measured flux by an unweighted average of approximately 1.45\%, 0.91\%, and 1.19\% of the {\tt SPIRECAL\_13\_1} flux, for the 250 \mum, 350 \mum, and 500 \mum~channels.  The shape of the histogram distribution is very close to Gaussian in each case, however the 250 \mum~channel shows a slight positive skewness.  The main updates in the calibration and data reduction pipeline are improved absolute flux calibrations of Neptune, and a better algorithm in destriping the data and removal of image artifacts.

In the second column of Figure \ref{fig:SpireCalComparison} we show the histogram of measured fluxes between {\tt SPIRECAL\_13\_1}, and the latest version {\tt SPIRECAL\_14\_2}.  The only change was an update to the absolute flux calibration of the instrument, which resulted in an even smaller change in the average flux: 0.24\%, --0.19\%, and 0.25\% for the 250 \mum, 350\mum, and 500 \mum~channels respectively.  SPIRE maps reduced using the two previous calibration versions are available upon request.

\ifnum\Mode=0
\placefigure{fig:SpireCalComparison}
\else
\begin{figure*}[h*]
\figurenum{8}
\label{fig:SpireCalComparison}
\center
\vskip -1.4in
\includegraphics[scale=0.85,angle=0]{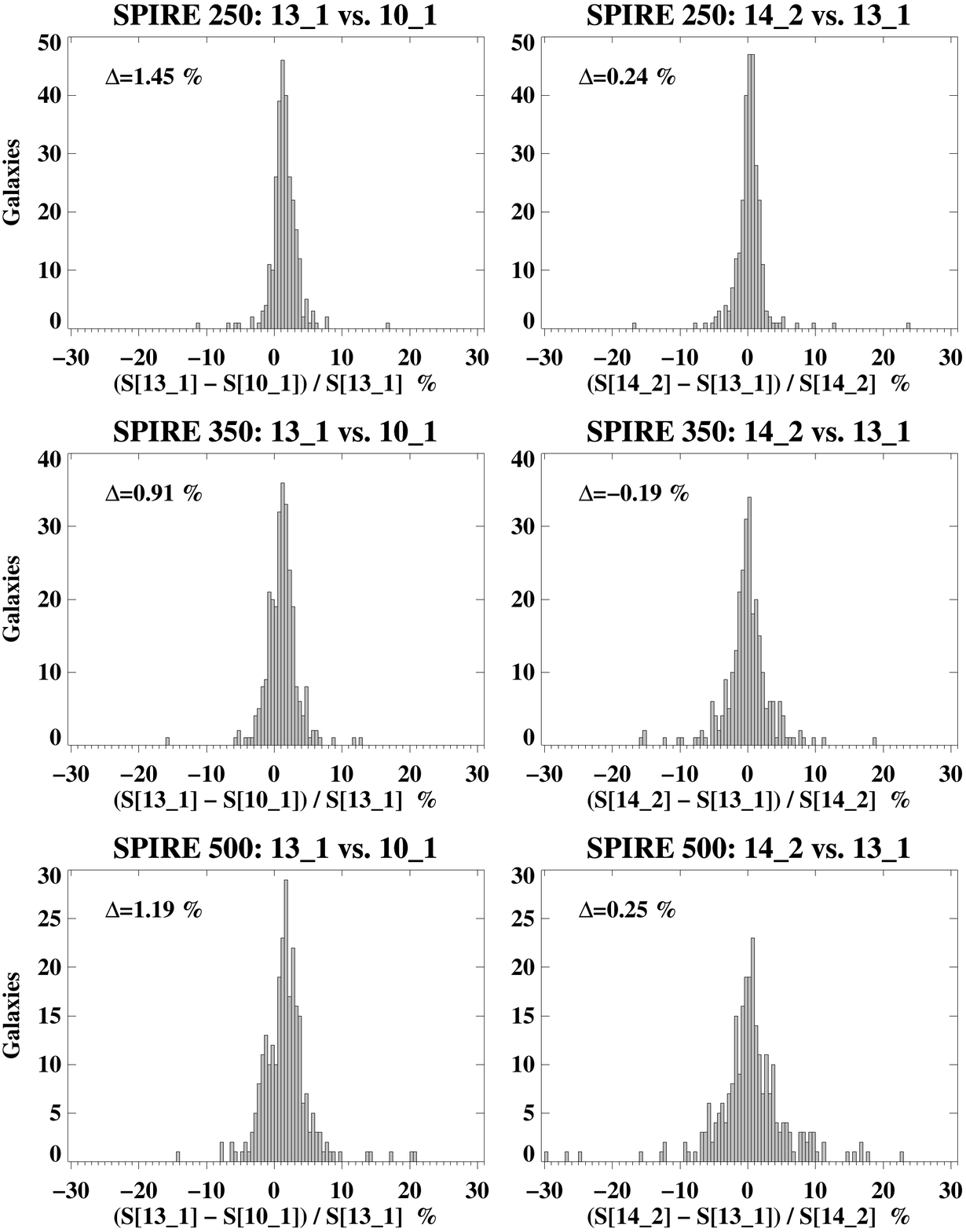}
\caption{Histogram plots comparing the percent change in flux between {\tt SPIRECAL\_10\_1} vs.~{\tt SPIRECAL\_13\_1} in the first column, and {\tt SPIRECAL\_13\_1} vs.~{\tt SPIRECAL\_14\_2} in the second column.  The values in each panel represent the unweighted average percent change between each calibration version.}
\end{figure*}
\fi

\subsection{Caveats}
In this section we detail several cautionary notes on using the data presented in this paper.

\subsubsection{Color Corrections}
By convention both of the PACS and SPIRE data reduction packages consider a flux calibration of the form $\nu F_\nu=\mathrm{constant}$ (\ie a spectral index of $-1$).  Since the {\em Herschel} photometry for the GOALS sample covers a wide range of wavelengths, and therefore different parts of the galaxy's SED, the color correction factor changes as a function of wavelength, as well as weaker dependence on infrared luminosity (due to a change in the dust temperature).  This is because the effective beam area of each instrument changes slightly for different spectral indices.  For PACS the color correction factors are listed on the NHSC website\footnote{\tt{\url{https://nhscsci.ipac.caltech.edu/pacs/docs/
PACS\_photometer\_colorcorrectionfactors.txt}}}, and are applied to the fluxes by {\em dividing} the factor for the appropriate power law exponent.  The SPIRE color correction factors are listed in the online SPIRE data reduction guide\footnote{\tt{\url{http://herschel.esac.esa.int/hcss-doc-14.0/print/
spire\_drg/spire\_drg.pdf}}} in Table 6.16 and are to be {\em multiplied}.

For this paper, we have decided to forego applying a color correction for both PACS and SPIRE fluxes.  This would otherwise require a detailed analysis involving a multi-component SED fit for each galaxy to derive the spectral slope at each observed {\em Herschel} band, which is outside the current scope of this paper.  This decision was agreed upon for both the {\em Herschel} and {\em Spitzer} (Mazzarella \etal 2017) data for the GOALS sample.  Flux changes due to color corrections for PACS bands are up to $\sim$$3\%$, and for SPIRE bands up to $\sim$$6\%$ for extended sources, which is less than or equal to the absolute calibration uncertainty of both instruments.  However we note for point sources, the SPIRE color correction can be higher, which we estimate to be $\sim$$15\%$ for a spectral index of $\alpha=4$.  If a photometric precision of within a few percent is desired, we strongly recommend users of the {\em Herschel}-GOALS data to include color corrections to the aperture photometry presented in this paper.

\subsubsection{PACS Saturation Limits}
Since galaxies within GOALS sample are very bright in the far-infrared, there is a small chance that some of our images exhibit saturation issues in a few of our {\em Herschel} maps.  For the PACS photometer there are two types of saturations.  Hard saturation occurs when the signal after the readout electronics are outside the dynamic range of the analog-to-digital converter.  On the other hand soft saturation arises from saturation of the readout electronics itself.  Taking into account both effects, the point source saturation limits are 220 Jy, 510 Jy, and 1125 Jy for the 70 \mum, 100 \mum, and 160 \mum~passbands respectively.

Fortunately for our sample, the latter two passbands have saturation limits well above our maximum measured fluxes of 248 Jy and 301 Jy for the 100 \mum~and 160 \mum~channels.  For the 70 \mum~channel, the nearby galaxy F02401--0013 has a total measured flux of 290 Jy which is above the saturation limit, and F06107+7822 which has a flux of 205 Jy and is close to the saturation limit.  However both appear very extended at 70 \mum, and in checking the saturation masks in the time-ordered data cubes we found no significant number of pixels were masked due to saturation.

\subsubsection{Correlated Noise in PACS Data}
Nine of our PACS maps exhibit residual correlated noise resembling low-level ripples in both the scan and cross-scan directions for only the blue camera (70 \mum~and 100 \mum).  Of these maps three of them only have this effect on the edges of the map, and do not affect the photometry or map quality.  Unfortunately for the other six maps the current processing techniques in Jscanam, Unimap, and MADMap fail to remove it.  One example of this is the 100 \mum~map of F03316--3618.  However we emphasize that these are very low-level effects, and do not significantly affect the quality of the photometry\footnote{These image artifacts are taken into account when calculating the uncertainty in flux.}, which we estimate to be on the few percent level.  This was calculated by first placing ten random apertures on empty sky on each map, then measuring the standard deviation in the flux per pixel on the affected maps.  This is then multiplied by the number of pixels within the photometry aperture.

\section{Summary}
In this paper we have presented broad band {\em Herschel} imaging for the entire GOALS sample in Figure \ref{fig:atlas}.  Total system fluxes, and component fluxes (where possible) are also computed in all six {\em Herschel} bands in Tables \ref{tbl:TotalFluxes} and \ref{tbl:ComponentFluxes} respectively.  Particular care was taken in producing archival quality atlas maps using the best data reduction codes and algorithms available at the time.  The data presented here are thus far the highest resolution, most sensitive and comprehensive far-infrared imaging survey of the nearest luminous infrared galaxies.  For many of these objects, this paper presents the first imaging data and reliable photometry at wavelengths beyond $\sim$200 \mum~in the submillimeter regime.

1) All 201 GOALS objects were detected in all three {\em Herschel} PACS (70, 100, and 160 \mum) and all three SPIRE (250, 350, 500 \mum) bands.  The FOV of the PACS and SPIRE images are sufficient and sensitive enough to detect the full extent of the far-infrared emission for even the widest pair separations.  Only two GOALS systems have full SPIRE coverage but lack PACS coverage of a third distant component (NGC 2385 in F07256+3355, and NGC 7769 in F23488+1949).  In addition, four galaxies observed outside of our {\em Herschel} program lack 100 \mum~data since they were not observed by those programs.

2) The image quality of the data are superb and were cleaned using the most up to date reduction routines and calibration files from the {\em Herschel} Science Center.  None of the images suffer from any saturation effects, major striping, or other image quality issues that may arise from scan-based observations.  Aperture corrections were applied only to point sources, while no color corrections were applied to any objects.  Furthermore the SPIRE 350 \mum~and 500 \mum~maps were zero-point corrected using data from the {\em Planck} observations.

3) The resolution is sufficient to resolve individual components of many pairs and interacting/merging systems in our sample, particularly at the shorter wavelengths where the PACS 70 \mum~FWHM band has a beamsize of $5\arcs6$. On the other hand wider pairs can still be resolved even at the longer wavelength SPIRE bands.

4) Comparing our PACS 70 and 100 \mum~fluxes to the legacy {\em IRAS} 60 and 100 \mum~measurements respectively, we found an excellent agreement (to within error) across our flux range as well as object morphologies ranging from point sources to extended systems.

5) The PACS 70 \mum~and 160 \mum~data within this paper supersede the reported fluxes and maps from the MIPS instrument on {\em Spitzer} (see Mazzarella \etal 2017, in prep.) due to the better sensitivity, resolution, and lack of image artifacts in the {\em Herschel} data.

In conjunction with datasets from other infrared telescopes (\ie {\em Spitzer}, WISE), the {\em Herschel} data from this paper will allow us for the first time to construct accurate spectral energy distributions in the infrared ($\sim$3--500 \mum) for the entire GOALS sample, which will be presented in several forthcoming papers.  The FITS files for the image mosaics constructed and presented in this atlas are being made available in the Infrared Science Archive (IRSA)\footnote{\tt{\url{http://irsa.ipac.caltech.edu/data/Herschel/GOALS/}}}.  Metadata for the images are also being folded into the NASA/IPAC Extragalactic Database (NED)\footnote{\tt{\url{http://ned.ipac.caltech.edu/}}} to simplify searches in context with other data in NED, including links to the FITS files at IRSA.

\section{Acknowledgments} 
J. Chu gratefully acknowledges Laurie Chu for proofreading the manuscript, and Thomas Shimizu for discussions on reducing the {\em Herschel} data.  D.S.~acknowledges the hospitality of the Aspen Center for Physics, which is supported by the National Science Foundation Grant No.~PHY-1066293.  D.S.~and K.L.~also acknowledge the Distinguished Visitor Program at the Research School for Astronomy and Astrophysics, Australian National University for their generous support while they were in residence at the Mount Stromlo Observatory, Weston Creek, NSW.  J.B., J.C., K.L. and D.S.~gratefully acknowledge funding support from NASA grant NNX11AB02G.  G.C.P. was supported by a FONDECYT Postdoctoral Fellowship (No.\ 3150361).  Support for this research was provided by NASA through a GO Cycle 1 award issued by JPL/Caltech.  We thank the Observer Support group of the NASA {\em Herschel} Science Center for patiently handling revisions and refinements of our AORs before execution of the observations, and for their expert assistance in reducing the data.  This paper has used data from the {\em Planck} mission, which is a project of the European Space Agency in cooperation with the scientific community.  ESA led the project, developed the satellite, integrated the payload into it, and launched and operated the satellite.  This research has made extensive use of the NASA/IPAC Extragalactic Database (NED) which is operated by the Jet Propulsion Laboratory, California Institute of Technology, under contract with the National Aeronautics and Space Administration.  We thank the anonymous referee whose comments helped us further improve our manuscript.

Facilities: {\em Herschel} (PACS), {\em Herschel} (SPIRE), {\em Planck} (HFI).

\clearpage

\bibliography{astro_refs}

\end{document}